\documentclass[aps,prb,showpacs,reprint]{revtex4-1}

\usepackage{amsmath}
\usepackage{graphicx}

\begin{document}

\title{Capture numbers and island size distributions\\ in models of
   submonolayer surface growth}

\author{Martin K\"orner}
\author{Mario Einax}
\email{mario.einax@uni-osnabrueck.de}
\author{Philipp Maass}
\email{philipp.maass@uni-osnabrueck.de}
\homepage{http://www.statphys.uni-osnabrueck.de}
\affiliation{Fachbereich Physik, Universit\"at Osnabr\"uck,
Barbarastra{\ss}e 7, 49076 Osnabr\"uck, Germany}

\date{May 27, 2012}

\begin{abstract}
  The capture numbers entering the rate equations (RE) for
  submonolayer film growth are determined from extensive kinetic Monte
  Carlo (KMC) simulations for simple representative growth models
  yielding point, compact, and fractal island morphologies. The full
  dependence of the capture numbers $\sigma_s(\Theta,\Gamma)$ on
  island size $s$, and on both the coverage $\Theta$ and the
  $\Gamma=D/F$ ratio between the adatom diffusion coefficient $D$ and
  deposition rate $F$ is determined. Based on this information, the RE
  are solved to give the RE island size distribution (RE-ISD), as
  quantified by the number $n_s(\Theta,\Gamma)$ of islands of size $s$
  per unit area. The RE-ISDs are shown to agree well with the
  corresponding KMC-ISDs for all island morphologies. For compact
  morphologies, however, this agreement is only present for coverages
  smaller than $\Theta\simeq5\%$ due to a significantly increased
  coalescence rate compared to fractal morphologies. As found earlier,
  the scaled KMC-ISDs $n_s\bar s^2 /\Theta$ as a function of scaled
  island size $x=s/\bar s$ approach, for fixed $\Theta$, a limiting
  curve $f_\infty(x,\Theta)$ for $\Gamma\to\infty$. Our findings provide
  evidence that the limiting curve is independent of $\Theta$ for
  point islands, while the results for compact and fractal island
  morphologies indicate a dependence on $\Theta$.
 \end{abstract}



\maketitle

\section{Introduction}
\label{sec:introduction}

The kinetics of submonolayer nucleation and island growth during the
initial stage of epitaxial thin film growth has been studied
intensively both experimentally and theoretically for more than three
decades (for reviews, see Refs.~\onlinecite{Brune:1998,
  Michely/Krug:2004, Evans/etal:2006, Dieterich/etal:2008}, and
references therein). Important aspects of the growth kinetics in the
submonolayer growth regime can be described by the rate equations (RE)
approach.\cite{Venables:1973} This approach has proven to be very
valuable in inorganic thin film growth. Interestingly, many of the
theoretical concepts developed for thin film growth kinetics of
inorganic materials, have shown recently to be very valuable also for
applications in organic thin film growth.\cite{Hlawacek/etal:2008,
  Loske/etal:2010, Zangwill/Vedensky:2011, Potocar/etal:2011,
  Koerner/etal:2011} This is due to the fact that these concepts often
are not specifically referring to particular materials. Instead, they
take into account the key mechanisms involved in the complex interplay
of deposition, evaporation, diffusion, aggregation and dissociation
from a general viewpoint.

Parameters entering the RE are the capture numbers
$\sigma_s(\Theta,\Gamma)$, which describe the strength of islands of
size $s$ to capture adatoms at a coverage $\Theta$ and ratio
$D/F\equiv\Gamma$ of the adatom diffusion coefficient $D$ and
deposition flux $F$.  The dependence of the capture numbers on $s$ has
been studied for various $\Gamma$ but only for one or a few $\Theta$
values. In this work we present a systematic study of the full
dependence on both $\Theta$ and $\Gamma$ for different types of island
morphologies and the case, where detachments of atoms from islands can
be neglected, corresponding to a critical nucleus of size $i=1$.  This
is motivated by the following questions, which have not been
thoroughly answered yet:
\begin{itemize}

\item[(1)] If the $\sigma_s(\Theta,\Gamma)$ are known, do the RE then
  predict correctly the number density $n_s(\Theta,\Gamma)$ of islands
  of size $s$, that means the island size distribution (ISD)?  This
  question indeed was earlier posed by Ratsch and Venables
  \cite{Ratsch/Venables:2003} as well as Evans {\it et al.}
  \cite{Evans/etal:2006} The answer to this question is not obvious,
  since the RE with known capture numbers $\sigma_s(\Theta,\Gamma)$
  neglect many-particle correlation effects,\cite{comm:many-particle}
  spatial fluctuations in shapes and capture zones of islands, and
  coalescence events that, despite rare in the early-stage growth, can
  have a significant influence.\cite{Koerner/etal:2010}

\item[(2)] Is there a simple functional form of the
  $\sigma_s(\Theta,\Gamma)$, in particular, is there a scaling of
  these capture numbers with respect to an effective capture length as
  suggested by a self-consistent treatment \cite{Bales/Chrzan:1994,
    Bales/Zangwill:1997} based on the RE?  Do the $\sigma_s$, when
  scaled with respect to their mean $\bar\sigma$, depend for large
  $\Gamma$ on the scaled island size $s/\bar s$ only, as suggested by
  Bartelt and Evans \cite{Bartelt/Evans:1996}?

\end{itemize}

In previous studies it has been found that the scaled ISD $\bar s^2
n_s/\Theta$ as a function of scaled island size $x=s/\bar s$
approaches, for fixed coverage $\Theta$, a scaling function $f(x)$ for
large $\Gamma$. Early simulations suggested that $f(x)$ is independent
of $\Theta$ and moreover not sensitive to the island
morphology. However, later results showed \cite{Bartelt/Evans:1996,
  Evans/etal:2006} that the morphology has an influence on the form of
$f(x)$. In fact, one would expect the scaling function $f(x)$ to
become independent of $\Theta$ if the RE with known capture numbers
correctly predict the ISD, and if the scaled capture numbers
$\sigma/\bar\sigma$ as a function of $s/\bar s$ become independent of
$\Theta$ for large $\Gamma$. Under these assumptions, an explicit
relation was proposed by Bartelt and Evans,\cite{Bartelt/Evans:1996,
  Evans/Bartelt:2001} which connects the scaling function of the
capture numbers with the scaling function of the ISD. We hence address
the following further questions:

\begin{itemize}

\item[(3)] Is the scaling function $f(x)$ independent of $\Theta$ for
  large $\Gamma$? What is the influence of the island morphology? Can
  the relation between the scaled ISD $f(x)$ and the scaling function
  for the capture numbers be confirmed?
\end{itemize}

The RE treatment is based on a coupled set of simple rate equations
describing the time evolution of the adatom density $n_1$ and the
number density $n_s$ of islands with size $s\geq2$, if spatial
correlations among islands during growth are neglected.  Taking into
account direct impingement of arriving atoms at the border of islands,
the RE for the case $i=1$ read
\begin{align}
\label{eq:n1}
\frac{1}{F}\frac{\mathrm d n_1}{\mathrm d t} =& (1-\Theta) - 2
\Gamma \sigma_1
n_1^2 - \Gamma n_1 \sum_{s>1} \sigma_s n_s\nonumber\\
&{}- 2 \kappa_1 n_1 - \sum_{s>1} \kappa_s
n_s \, \\
\frac{1}{F}\frac{\mathrm d n_s}{\mathrm d t} =& \Gamma n_1 \left(
  \sigma_{s-1} n_{s-1} - \sigma_s
  n_s \right)\nonumber\\
&{}+ \kappa_{s-1} n_{s-1} - \kappa_s n_s \,,\quad s=2,3,\ldots
\label{eq:ns}
\end{align}
These equations refer to the pre-coalescence regime where only adatoms
are mobile and it is presumed that re-evaporation of atoms and atom
movements between the first and second layer can be
disregarded. Moreover, adatoms arriving on top of an island are not
counted, i.e.\ $s$ in a strict sense refers to the number of substrate
sites covered by an island (or the island area).  The coverage
$\Theta$ entering Eq.~(\ref{eq:n1}) is given by $\Theta=\sum_{s\geq1}
s n_s=1-\exp(-Ft)$ and takes into account that adatoms are generated
by deposition into the uncovered substrate area (as common in the
literature in this field, we set the length unit equal to the the size
of the substrate lattice unit). The terms with $\sigma_1(\Theta)$ and
$\kappa_1(\Theta)$ describe the nucleation of dimers due to attachment
of two adatoms by diffusion and due to direct impingement,
respectively. The term $\propto n_1\sigma_s n_s$ describes the
attachment of adatoms to islands of size $s>1$, and the term
$\propto\kappa_s n_s$ the direct impingement of deposited atoms to
boundaries of islands with size $s$. For the idealized point island
model, $s$ refers to the total number of atoms that arrived at a
point, and $(1-\Theta)$ in Eq.~(\ref{eq:n1}) is replaced by one (no
covered substrate area). For a unified discussion of capture numbers
and the ISD we formally set $\Theta=Ft$ for the point island model.

Introducing the total number density of stable islands $N$ and the
average capture number $\overline{\sigma}$,
\begin{align}
\label{eq:number_of_stable_islands-sigma}
N &=\sum\limits_{s>1} n_s, \quad \overline{\sigma} = \frac{1}{N}
\sum_{s>1} \sigma_s n_s \, ,
\end{align}
a reduced set of equations for $n_1(\Theta)$ and $N(\Theta)$ can be
derived from Eqs.~(\ref{eq:n1}) and (\ref{eq:ns}) within the RE
treatment.  These equations predict the scaling relation $N \propto
\Gamma^{-\chi}$ with the scaling exponent
$\chi=1/3$.\cite{Venables/etal:1984,Venables:1994} This relation has
been successfully validated by several growth experiments in the past
and applied to extract adatom diffusion barriers and binding energies
in metal epitaxy. A discussion of many of these experiments can be
found in Ref.~\onlinecite{Michely/Krug:2004}.  Recently, the relation
has also been applied in organic thin film growth
.\cite{Loske/etal:2010,Hlawacek/etal:2008} An extended RE approach
for multicomponent adsorbates
\cite{Einax/etal:2007,Dieterich/etal:2008} was recently suggested to
determine binding energies between unlike atoms from island density
data.\cite{Einax/etal:2009}

More detailed information on the growth kinetics is contained in the
ISD. If the full dependence of the ISD $n_s(\Theta,\Gamma)$ on
$\Theta$ and $\Gamma$ is mediated by the mean island size $\bar
s(\Theta,\Gamma)$, the ISD should obey the following scaling form, as
first suggested by Vicsek and Family,\cite{Vicsek/Family:1984}
\begin{align}
\label{eq:scaling_form}
\frac{{\overline s}^2 (\Theta, \Gamma)}{\Theta} n_s (\Theta,\Gamma)=
f\left(\frac{s}{\bar s(\Theta,\Gamma)}\right) \;\;.
\end{align}
Here the scaling function $f(x)$ must fulfill the conditions
$\int_0^{\infty} f(x) dx = \int_0^\infty x f(x) dx =1$.  The scaling
behavior was found to give a good effective description for large
$\Gamma$. More precisely, the curves ${\overline s}^2n_s/\Theta$ as a
function of $x=s/\bar s$ approach a limiting curve,
\cite{Vvedensky:2000}
\begin{align}
\label{eq:limiting_curve}
\lim_{\Gamma\to\infty} \frac{{\bar s}^2}{\Theta} n_{x\bar s}
(\Theta,\Gamma)= f_\infty(x,\Theta) \, .
\end{align}
Previous studies for a few fixed $\Theta$ values suggest that
$f_\infty(x,\Theta)$ is independent of $\Theta$.

An explicit expression for the scaling function $f(x)$ with shape
independent of $\Theta$ was suggested by Amar and
Family,\cite{Amar/Family:1995}
\begin{align}
\label{eq:empirical_form}
f(x) = C_{i} x^{i} \exp \left(- i a_{i} x^{1/a_{i}} \right) \, .
\end{align}
The parameters entering this scaling function depend on the size of
the critical nucleus $i$, which allows one to determine $i$ in
experiments.\cite{Loske/etal:2010, Potocar/etal:2011, Ruiz/etal:2003,
  Pomeroy/Brock:2006} Equation~(\ref{eq:empirical_form}) was believed
to be independent even of the morphology \cite{Amar/Family:1995}, but
this has later been questioned.\cite{Bartelt/Evans:1996,
  Evans/etal:2006}

Based on a continuum limit of the RE (\ref{eq:ns}) and scaling
assumptions for the capture numbers and a neglect of the
$\Theta$-dependence, an expression for the limiting curve
$f_\infty(x)$ was derived by Bartelt and Evans,
\cite{Bartelt/Evans:1996, Evans/Bartelt:2001}
\begin{align}
\label{eq:asympotic_form}
f_\infty(x) = f_\infty(0)\exp\left\{\int_0^{x} {\rm d}y
  \frac{(2z_\infty-1)-C_{\rm tot}'(y)}{C_{\rm tot}(y)-zy} \right\} \,,
\end{align}
where $z_\infty={\mathrm \partial(\ln \bar s)/\mathrm \partial(\ln
  \Theta)}$, and $C_{\rm tot}(x)$ is a linear combination of the
scaled capture numbers $C_\infty (x)=\sigma_s/\bar\sigma$ and scaled
direct capture areas $K_\infty (x)=\kappa_s/\bar\kappa$. The
$\infty$-subscript indicates that the large $\Gamma\to\infty$ limit
should be taken. As pointed out by Bartelt and Evans, $C_{\rm tot}(x)$
should be well approximated by the scaled capture numbers alone,
$C_{\rm tot}(x)\approx (1-\Theta)C_\infty(x)$. In Appendix~\ref{app:A}
we show that in fact it holds $C_{\rm tot}(x)\approx C_\infty(x)$. The
two conditions for $f_\infty(x)$ (normalization and first moment equal
to one) imply that $C_\infty (0)=(1-z)/f_\infty(0)$ and $\int_0^\infty
dx\,C_\infty(x)f_\infty(x)=1$. \cite{Evans/Bartelt:2001,Evans/etal:2006}

It is interesting to note that a semi-empirical form, which has a
structure similar to Eq.~(\ref{eq:empirical_form}) has been suggested
recently by Pimpinelli and Einstein \cite{Pimpinelli/Einstein:2007}
for the distribution of capture zones $A$ as identified by Voronoi
tessellation,
\begin{align}
\label{eq:Capture_zone_distribution}
P_\beta &= c_\beta a^\beta \exp(-d_\beta a^2) \, ,
\end{align}
where $a=A/\bar{A}$ is the rescaled capture zone with respect to the
mean $\bar{A}$ and $\beta=i+1$ (see also
Ref.~\onlinecite{Oliveira/Reis:2011}). This distribution corresponds
to a generalized Wigner surmise from random matrix theory. The
parameter $\beta=i+1$ and the functional form, however, are
controversially discussed.\cite{Li/etal:2010,
  Pimpinelli/Einstein:2010}

Besides this recent progress in predicting functional forms of capture
zone distributions, there are only a few studies so far
\cite{Gibou/etal:2001, Gibou/etal:2003, Koerner/etal:2010} that
address the problem whether an integration of the RE (\ref{eq:n1})
and (\ref{eq:ns}) can yield correctly the ISD for different cluster
morphologies in the pre-coalescence regime.  For an integration of the
RE a reliable determination of $\sigma_s(\Theta,\Gamma)$ is
needed. Four general approaches have been followed for this purpose:
(i) Within a self-consistent ansatz one can solve the diffusion field
around an island and derive determining equations for the capture
numbers by equating the attachment currents of the diffusion field and
the RE.  \cite{Bales/Chrzan:1994, Bales/Zangwill:1997} (ii) By
modeling the island growth with the level set method,
\cite{Gyure/etal:1998} one can analogously equate attachments
currents and determine the capture numbers.\cite{Gibou/etal:2001,
  Gibou/etal:2003} (iii) Balancing the deposition rate $FA_sn_s$ into
the mean capture zone $A_s$ of islands of size $s$ with the RE
expression $D\sigma_sn_1n_s$ for the attachment rate to these islands,
yields $\sigma_s\simeq A_s/\Gamma n_1$. This means that the capture
numbers can be approximately calculated from a determination of the
$A_s$, e.g.\ by Voronoi tessellation.\cite{Bartelt/etal:1999,
  Evans/Bartelt:2001, Bartelt/etal:1998a, Hannon/etal:1998} (iv) In
simulations, where the individual attachments are followed, the
capture numbers can be calculated from the mean number of attachments
$M_s$ to island of size $s$ during a time interval $\Delta t$ [see
Eq.~(\ref{eq:capture_numbers}) and the discussion in
Sec.~\ref{sec:II}].\cite{Bartelt/Evans:1996}

The paper is organized as follows. First we describe in
Sec.~\ref{sec:II} the method used to generate point, compact and
fractal island morphologies, and the method for determining the
capture numbers as function of island size and coverage. In
Sec.~\ref{sec:capture_numbers} we discuss the results for the capture
numbers and compare these with the prediction of the self-consistent
theory. In Sec.~\ref{sec:adatom_and_island_density} we analyze the
mean island and adatom densities for the different island morphologies
and discuss their prediction by the self-consistent RE and the RE
based on the capture numbers determined in the KMC simulations. In
Sec.~\ref{sec:ISD} we demonstrate that the ISD is successfully
predicted by the RE as long as coalescence events can be neglected.
These coalescence events are relevant already for small coverages
$\Theta\gtrsim0.05$ for compact morphologies, while they turn out to
be much less important for fractal morphologies.  The reason for these
differences are reduced coalescence rates for fractal island
morphologies because of a screening
effect.\cite{Brune:1998,Brune/etal:1999} Finally, we study in
Sec.~\ref{sec:Scaling_theory} the behavior of the scaled capture
numbers and scaled ISDs in the limit $\Gamma\to\infty$.

\section{Submonolayer growth: models,
morphologies and simulations}
\label{sec:II}
The KMC simulations are performed with a first reaction time Monte
Carlo algorithm \cite{Holubec/etal:2011, Gillespie:1977} on a square
lattice with $L\times L$ sites. In this algorithm, two times $\tau_F$
and $\tau_D$ are randomly generated from the exponential probability
density $\psi(\tau)=\gamma\exp(-\gamma\tau)$, where $\gamma=L^2F$ for
$\tau_F$, corresponding to a deposition process, and $\gamma=4DL^2n_1$
for $\tau_D$, corresponding to one of the possible diffusive jumps of
adatoms.  If $\tau_D<\tau_F$, the simulation time is incremented by
$\tau_D$ and one of the $L^2n_1$ adatoms is selected randomly and
moved to a randomly selected vacant nearest neighbor site. If
$\tau_F<\tau_D$, the simulation time is incremented by $\tau_F$ and
one of the $L^2$ sites is randomly chosen. If this site is vacant, an
additional adatom is deposited on this site, while, if the site is
occupied, no deposition takes place.

With respect to the formation of islands we consider three simple
growth models that are representative for the different types of
island morphologies in the case of $i=1$.  Fractal islands are
generated by applying ``hit and stick'' aggregation, that means an
adatom having another atom as nearest neighbor becomes
immobilized. Compact island morphologies are produced by letting
islands grow spirally into a quadratic form as in
Ref.~\onlinecite{Bartelt/Evans:1993}, meaning that each adatom
attaching to an island is displaced to the corresponding tip of the
spiral. Point island morphologies are generated by displacing an
adatom attaching to an island to the site representing the island,
while bookkeeping the total number of aggregated atoms for the island
size.

\begin{figure*}[t!]
\hfill\includegraphics[scale=0.22]{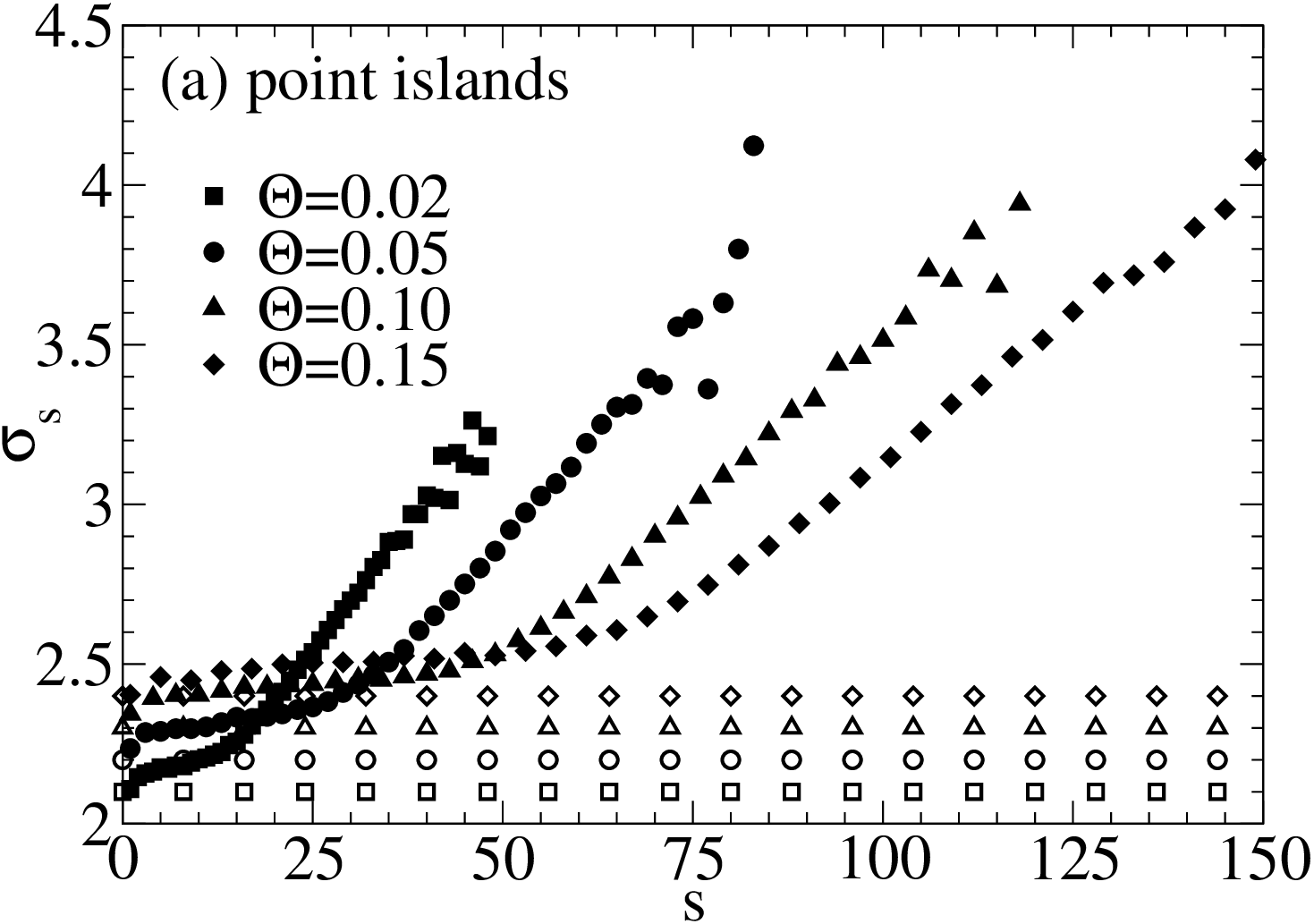}
\hfill\includegraphics[scale=0.22]{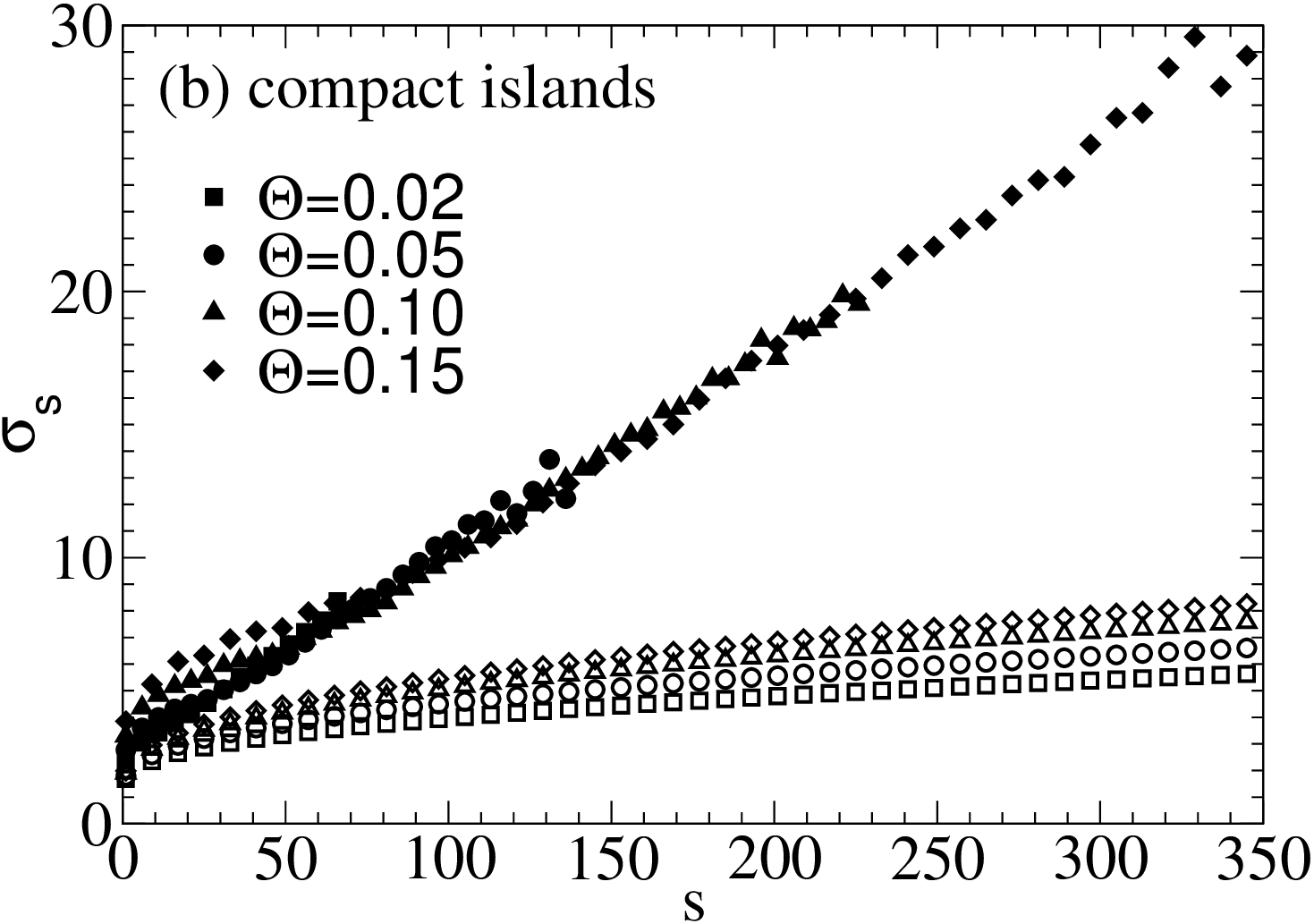}
\hfill\includegraphics[scale=0.22]{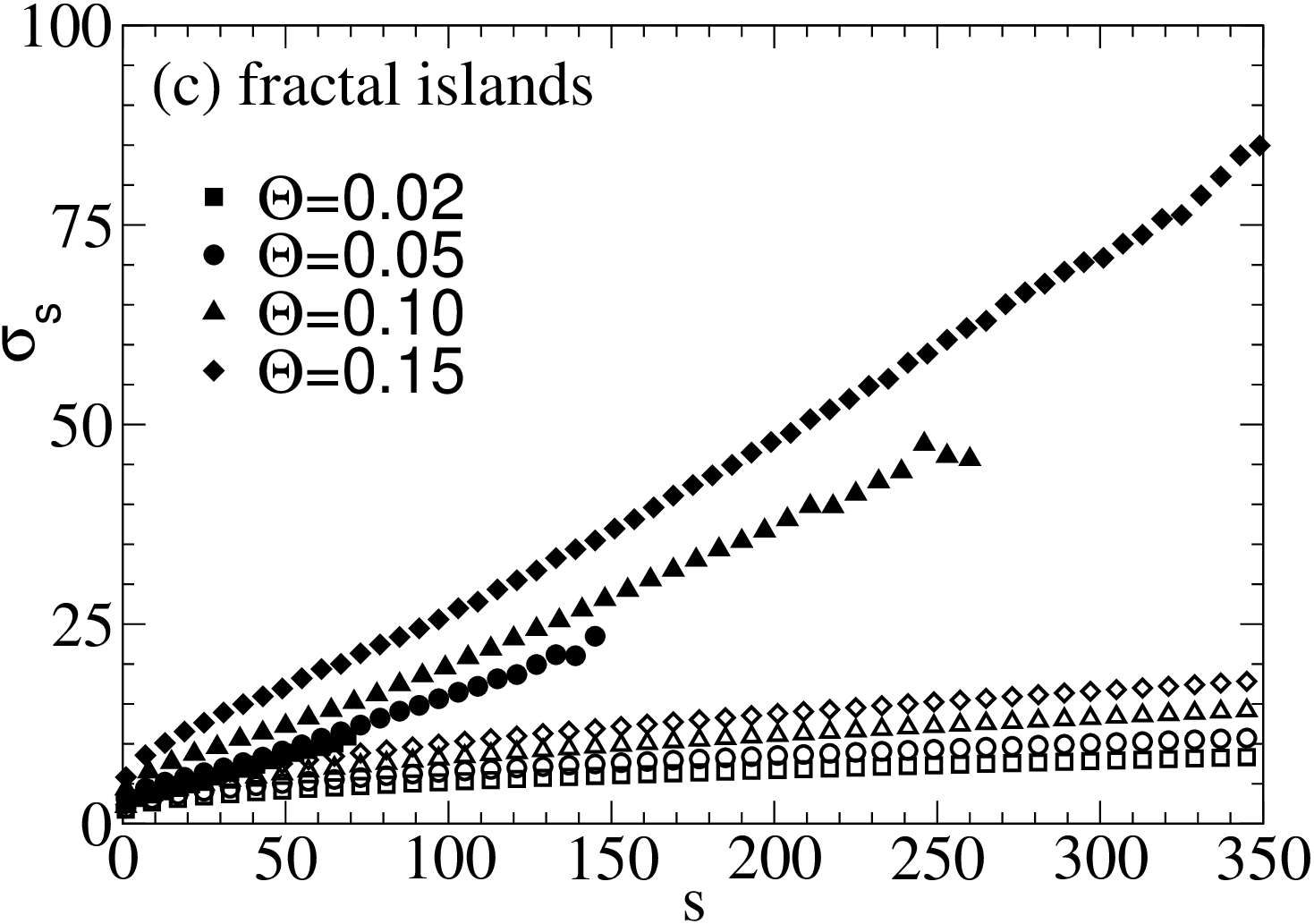}
\caption{Capture numbers as a function of $s$ for $\Gamma=10^7$ and
  four different fixed coverages for the models representing (a)
  point, (b) compact, and (c) fractal island morphologies. The filled
  symbols refer to the $\sigma_s$ obtained from the KMC simulations
  and the open symbols to the results of the self-consistent theory
  according to Eqs.~(\ref{eq:sigma_SC}) (with the $R_s$ taken from the
  simulations, see text and Fig.~\ref{fig:rs}). \vspace*{3ex} }
 \label{fig:ch3-sigma}
\end{figure*}

\begin{figure*}[t!]
\includegraphics[width=0.32\textwidth]{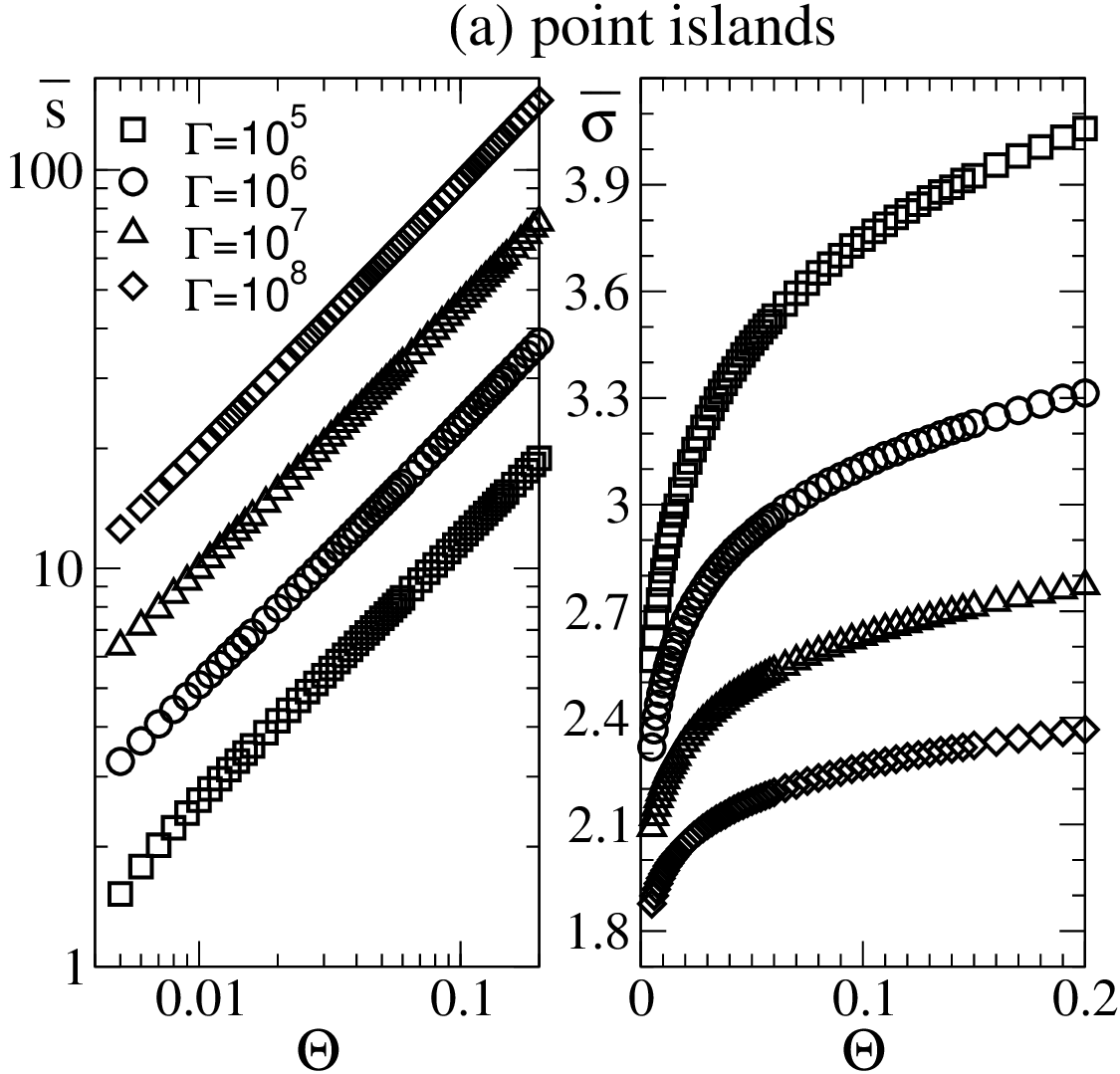}
\hspace{0.01\textwidth}\includegraphics[width=0.32\textwidth]{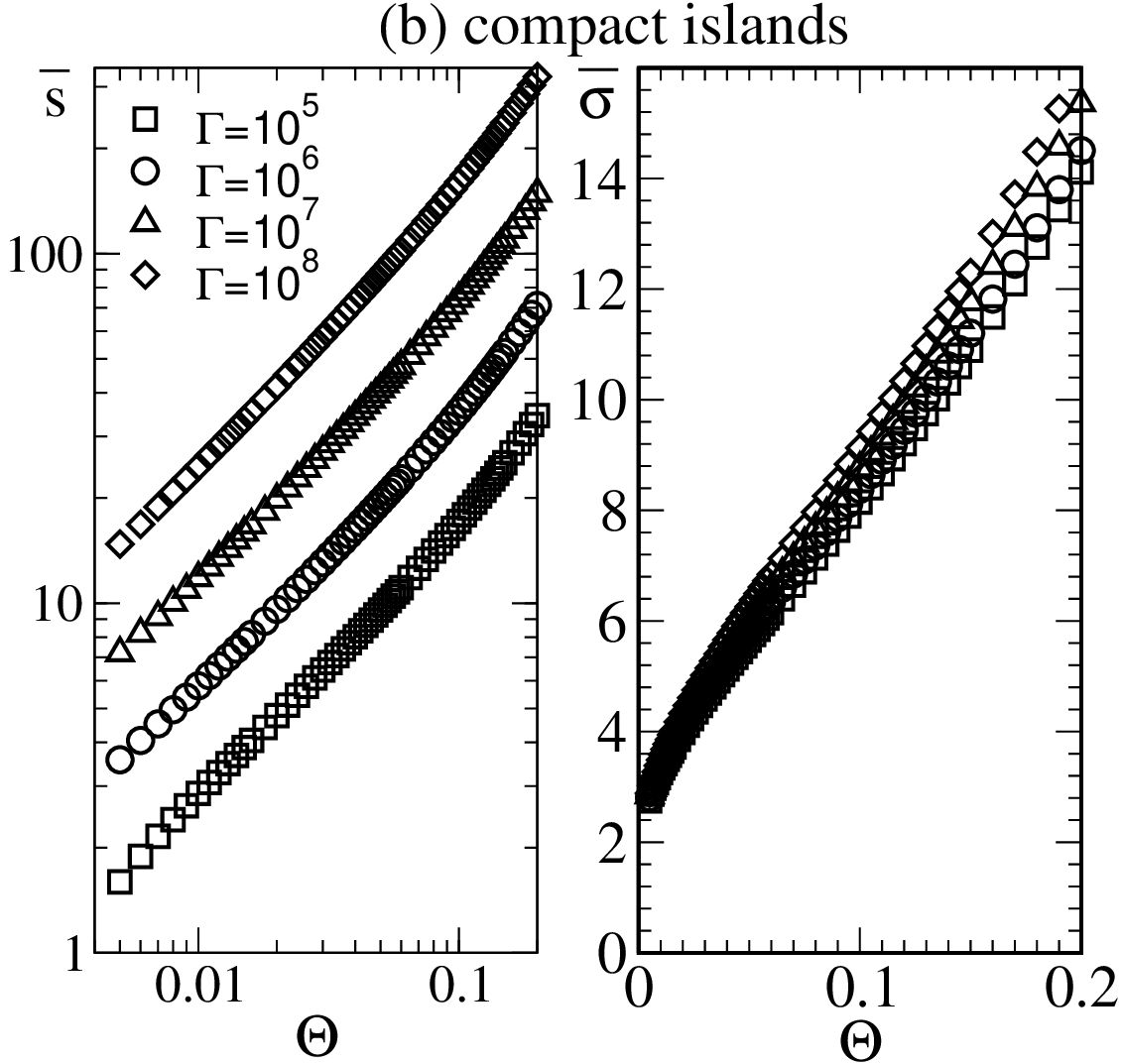}
\hspace{0.01\textwidth}\includegraphics[width=0.32\textwidth]{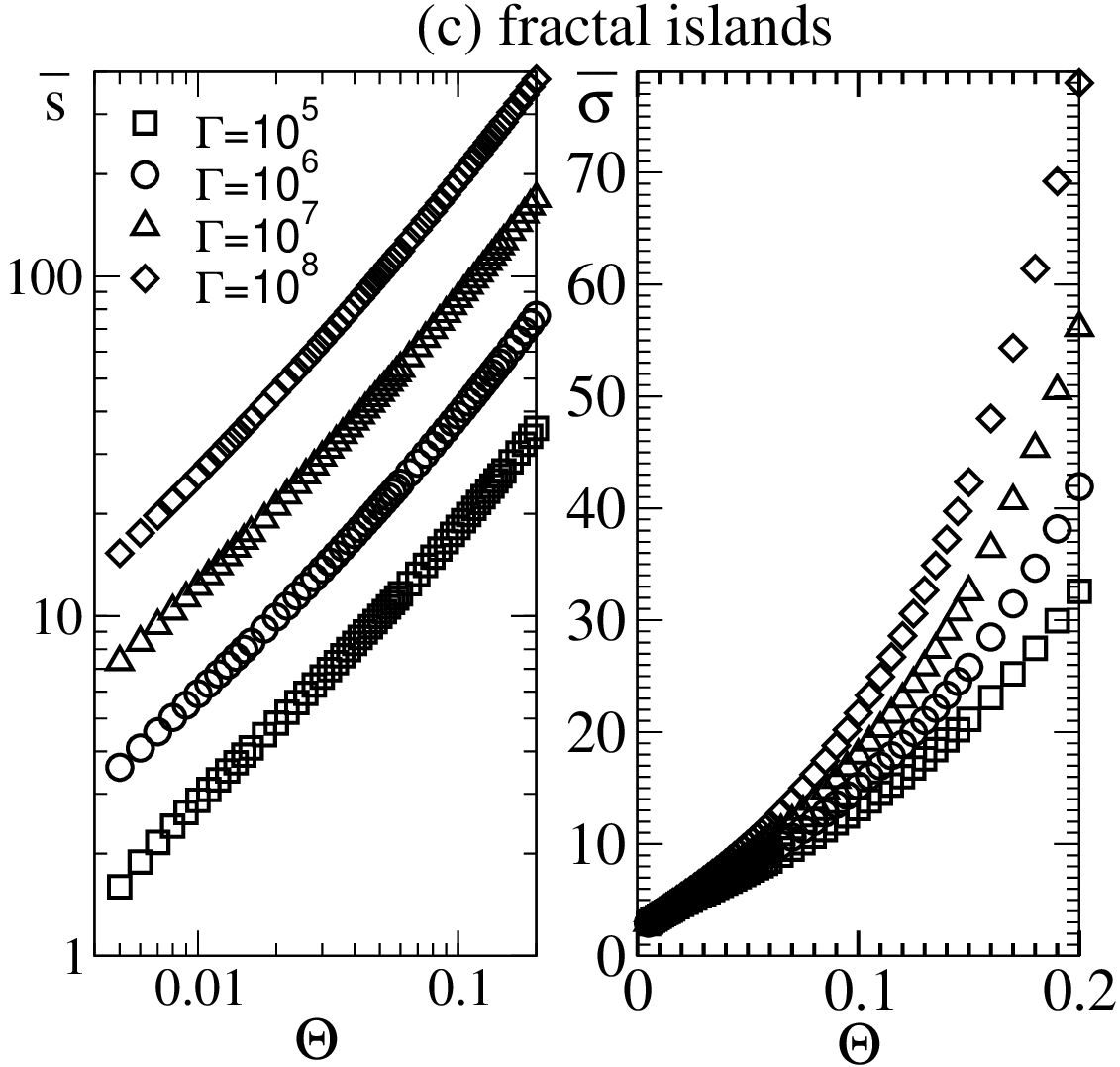}
\caption{Mean capture number $\bar\sigma$ and mean island size $\bar
  s$ as a function of $\Theta$ for the four simulated $\Gamma$ values
  and the models representing point, compact and fractal island
  morphologies.}
 \label{fig:mean-s-sigma}
\end{figure*}

To calculate the capture numbers $\sigma_s$ at the coverage $\Theta$,
we use the following procedure which is based on the method outlined
in Ref.~\onlinecite{Bartelt/Evans:1996}: Each simulation run is
stopped at coverage $\Theta$ and the number densities $n_s=N_s/L^2$,
$s=1,2,\ldots$ are determined, where $N_s$ are the numbers of monomers
($s=1$) and islands ($s>1$). Then the simulation is continued for a
time interval $\Delta t$ without deposition and the following
additional rules are implemented: (i) if an adatom is attaching to an
island of size $s>1$, a counter $M_s$ is incremented and the adatom
thereafter repositioned at a randomly selected site on the free
substrate area (i.e.\ a site which is neither covered nor a nearest
neighbor of a covered site); (ii) if two adatoms form a dimer, a
counter $M_1$ is incremented and the two adatoms thereafter are
repositioned randomly as described in (i). In this way a stationary
state is maintained at the coverage $\Theta$. The mean attachment rate
per unit area to islands of size $s$ is $M_s/(L^2\Delta t)$, and
equating this with the expression $D\sigma_sn_sn_1$ from the RE
(\ref{eq:n1},\ref{eq:ns}) yields
\begin{align}
\label{eq:capture_numbers}
\sigma_s=\frac{M_s}{L^2\Delta t D n_sn_1} && s=1,2,\ldots
\end{align}
Averaging over many simulation runs (configurations) gives
$\sigma_s(\Theta,\Gamma)$. The $\kappa_s$ are determined from the
lengths of the islands boundaries, which are simultaneously monitored
during the simulation and averaged for each size $s$.

The continuous-time Monte Carlo (KMC) simulations are performed on a
square lattice with periodic boundary conditions and $L\times
L=8000\times8000$ sites for four different $\Gamma=10^5$, $10^6$,
$10^7$, and $10^8$.  For each value of $\Theta$ an average over $10^8$
nucleation/attachment events was performed.

\section{Capture numbers}
\label{sec:capture_numbers}
The direct capture areas for point islands on a square lattice are
given by $\kappa_s=4$. For compact and fractal islands the $\kappa_s$
increase as $\sim\sqrt{s}$ and $\sim s$, respectively, and their
dependence on $\Theta$ and $\Gamma$ is very weak. 

Representative results for the capture numbers $\sigma_s$ are shown in
Fig.~\ref{fig:ch3-sigma} as a function of $s$ for fixed $\Gamma=10^7$
and four different coverages for the (a) point, (b) compact, and (c)
fractal island morphologies.  For the other simulated $\Gamma$ values,
a similar behavior was obtained. The mean $\bar\sigma(\Theta,\Gamma)$
[see Eq.~(\ref{eq:number_of_stable_islands-sigma})] as a function of
$\Theta$ for all simulated $\Gamma$ values is displayed in
Fig.~\ref{fig:mean-s-sigma}, together with the mean island size $\bar
s(\Theta,\Gamma)$. These functions are later used in
Sec.~\ref{sec:Scaling_theory} when investigating the scaled capture
numbers $\sigma_s$ in connection with the scaled island densities in
the limit $\Gamma\to\infty$.

A common feature for all morphologies in Fig.~\ref{fig:ch3-sigma} is a
linear increase of $\sigma_s$ with $s$ for large $s>\bar s$. It can be
understood \cite{Evans/etal:2006} from the proportionality of the
$\sigma_s$ to the mean capture zone areas $A_s$, and the fact that
large islands typically exhibit large $A_s$, which led to the stronger
growth of these islands.  Since a twice as large capture zone gives on
average rise to a twice as large island, it holds $A_s\sim s$ and
hence $\sigma_s\propto A_s\sim
s$. \cite{Bartelt/Evans:1996,Evans/etal:2006}

With respect to the dependence on the coverage $\Theta$, the
$\sigma_s$ in Fig.~\ref{fig:ch3-sigma} have a quite different behavior
for the three morphologies in the regime $s>\bar s$: While for the
point islands the $\sigma_s$ decrease with $\Theta$, they are almost
independent of $\Theta$ for the compact islands, and they increase
with $\Theta$ for the fractal islands. Main reason for these
differences is that for point islands the number density $N$ continues
to increase with $\Theta$ (that means time $t=\Theta/F$) due to
ongoing nucleation of new islands, while for compact and fractal
morphologies, $N$ tends to saturate for larger $\Theta$, with less
pronounced saturation in the compact case (see
Sec.~\ref{sec:adatom_and_island_density} below).  During the growth in
the point island model, a large capture zone surrounding a large
island is, compared to the other two morphologies, more frequently
destroyed by a nucleation event in this zone, and the
$A_s\propto\sigma_s$ thus decrease with $\Theta$ for fixed $s>\bar s$.
Due to the higher nucleation rate and the missing spatial extension of
islands in the point island model, the corresponding $\sigma_s$ are
much smaller than for the compact and fractal morphologies. The larger
island extension and the strong capture of adatoms by finger tips in
the case of fractal islands lead to about five times larger $\sigma_s$
in comparison to the compact islands.

The differences with respect to the $\Theta$ dependence are also
reflected in the behavior of $\bar\sigma(\Theta,\Gamma)$ in
Fig.~\ref{fig:mean-s-sigma}. In fact, when considering the scaled
capture numbers $\sigma_s/\bar\sigma$, the $\Theta$ dependence for
$s>\bar s$ becomes qualitatively the same for all morphologies
(increase of $\sigma_s/\bar\sigma$ with $\Theta$, see
Sec.~\ref{sec:Scaling_theory} below).  For small $s<\bar s$, the
curves in Fig.~\ref{fig:ch3-sigma} show a nonlinear dependence of
$\sigma_s$ on $s$ for all morphologies.\cite{Bartelt/etal:1998a,
  Bartelt/etal:1998b, Amar/etal:2001} By combining the linear function
for large $s$ with a polynomial at small $s$, we fitted the results
for $\sigma_s$ for all simulated $\Theta$ and $\Gamma$ values.  These
fits, together with corresponding fits for the $\kappa_s$, were used
to integrate the RE (\ref{eq:n1}) and (\ref{eq:ns}).

The mean island size $\bar s$ in Fig.~\ref{fig:mean-s-sigma}
reproduces the behavior seen in many earlier
studies. \cite{Evans/etal:2006} In the point island model the straight
lines in the double logarithmic representation are in agreement with
$\bar s\sim\Theta^z$ with $z={2/3}$ as predicted by a scaling analysis
of the reduced RE.\cite{Bartelt/Evans:1992, Evans/etal:2006,
  Dieterich/etal:2008-2} In the case of the compact and fractal island
morphologies, the slope $z(\Theta)=\partial
\ln s(\Theta,\Gamma)/\partial\ln\Theta$ increases with $\Theta$ and
approaches $z\simeq1$ for both island morphologies. This is consistent
with a saturation ($\Theta$-independence) of the island density for
large $\Theta$ in the pre-coalescence regime, $\bar s\sim\Theta/N\sim
\Theta$.

\begin{figure}[t!]
\includegraphics[scale=0.25]{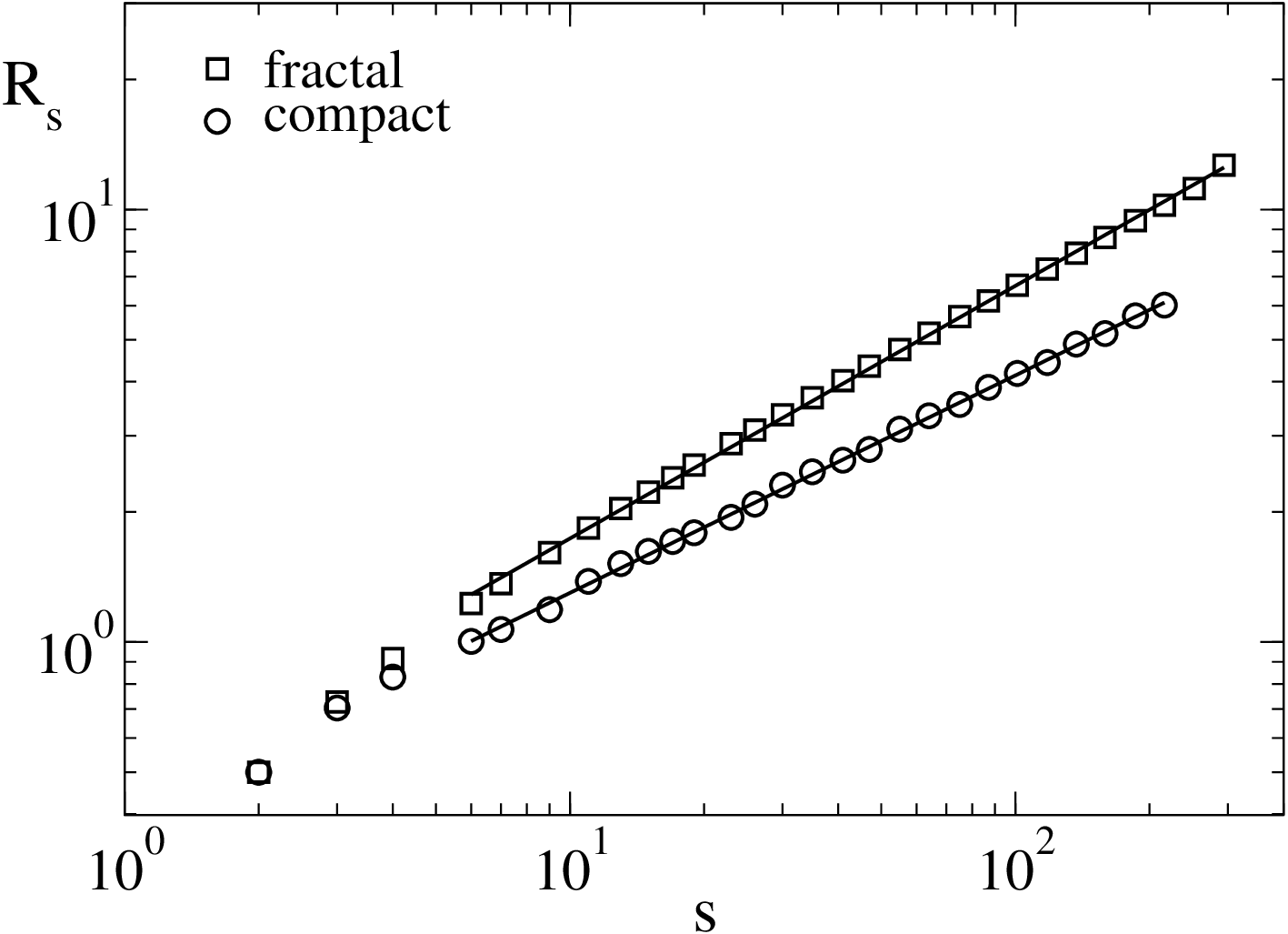} 
\caption{Mean radii of gyration $R_s$ of islands of size $s$ for the
  models representing compact and fractal island morphologies. The
  straight lines in the double-logarithmic plot indicate the
  power-law behavior for large $s$.}
\label{fig:rs}
\end{figure}

In the self-consistent theory \cite{Bales/Chrzan:1994} the capture
numbers are given by
\begin{subequations}
\label{eq:sigma_SC}
\begin{align}
\sigma_s^{\rm sc}&=
\frac{2\pi}{(1-\Theta)}\frac{R_s}{\xi}
\frac{{K}_1 \left(R_s/\xi \right)}{{K}_0 \left(R_s/\xi\right)}\,,
\label{eq:sigma_SCa}\\[1ex]
\xi^{-2} &= 2 \sigma_1^{\rm sc} n_1 + \sum_{s\geq 2} \sigma_s^{\rm sc} n_s \, ,
\label{eq:sigma_SCb}
\end{align}
\end{subequations}
where $R_s$ is the effective radius of an island of size $s$, $K_0$
and $K_1$ are the modified Bessel functions of order zero and one,
respectively, and $\xi$ is the adatom capture length (mean linear size
of depletion zone around an island). The factor $(1-\Theta)$ in
Eq.~(\ref{eq:sigma_SC}), which was not given in the original
derivation in Ref.~\onlinecite{Bales/Chrzan:1994} arises from the fact
that the adatom current to a (circular) island of size $s$ is $2\pi
R_sD\,\partial_r\tilde n_1(r)|_{r=R_s}$, where $\tilde
n_1(r)=n_1(r)/(1-\Theta)$ is the adatom density with respect to the
free (uncovered) surface area, and $n_1(r)$ is the local form
corresponding to the global mean value $n_1$ appearing in the RE
[see also Ref.~\onlinecite{Popescu/etal:2001a} for the additional
factor $(1-\Theta)$].  For $R_s \ll \xi$, $\sigma_s\sim
2\pi/[(1-\Theta)\ln(\xi/R_s)]$, and for $R_s \gg \xi$, $\sigma_s\sim
2\pi/(1-\Theta)(R_s/\xi)$.

\begin{figure}[t!]
\includegraphics[scale=0.25]{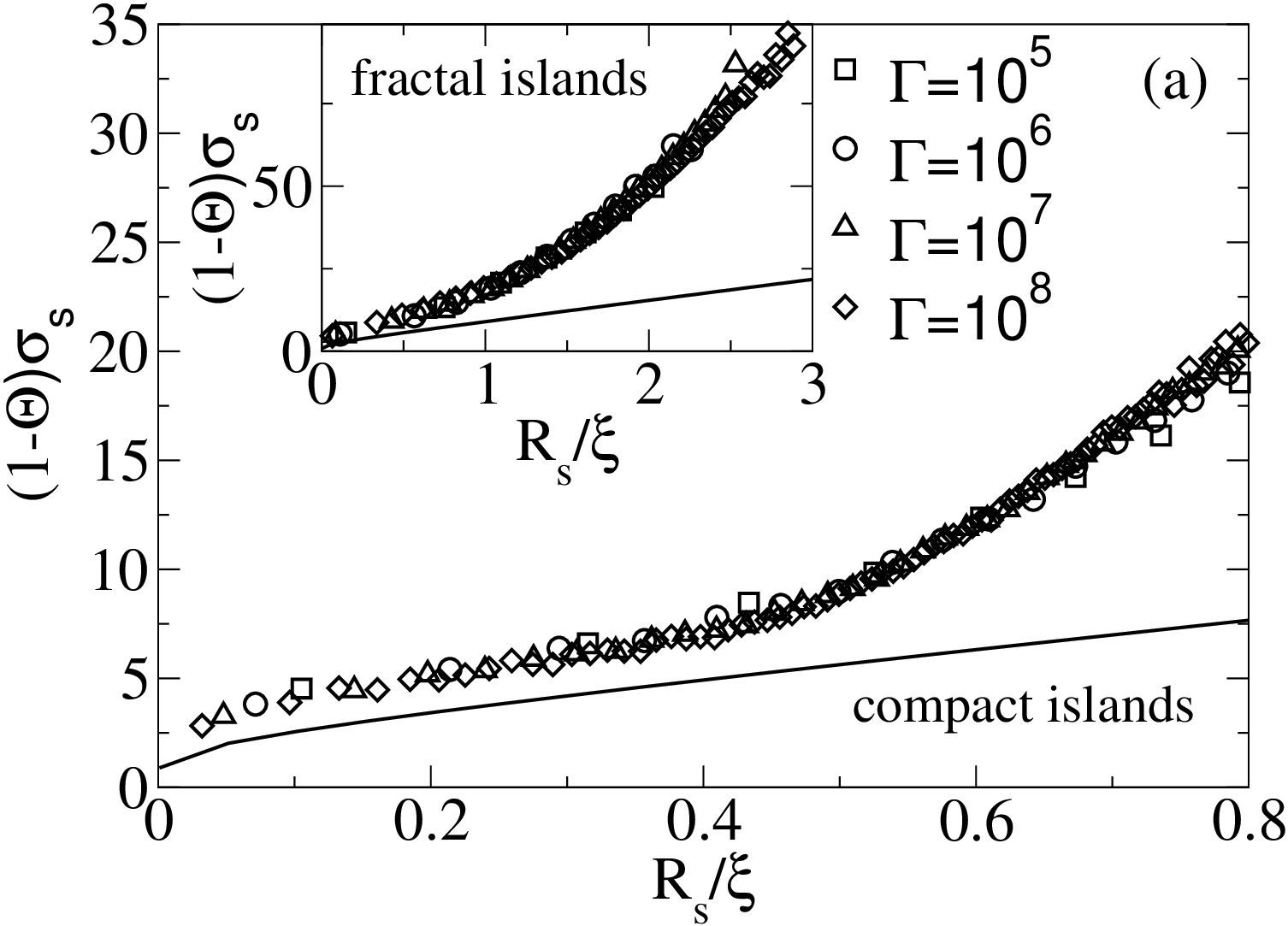}
\includegraphics[scale=0.25]{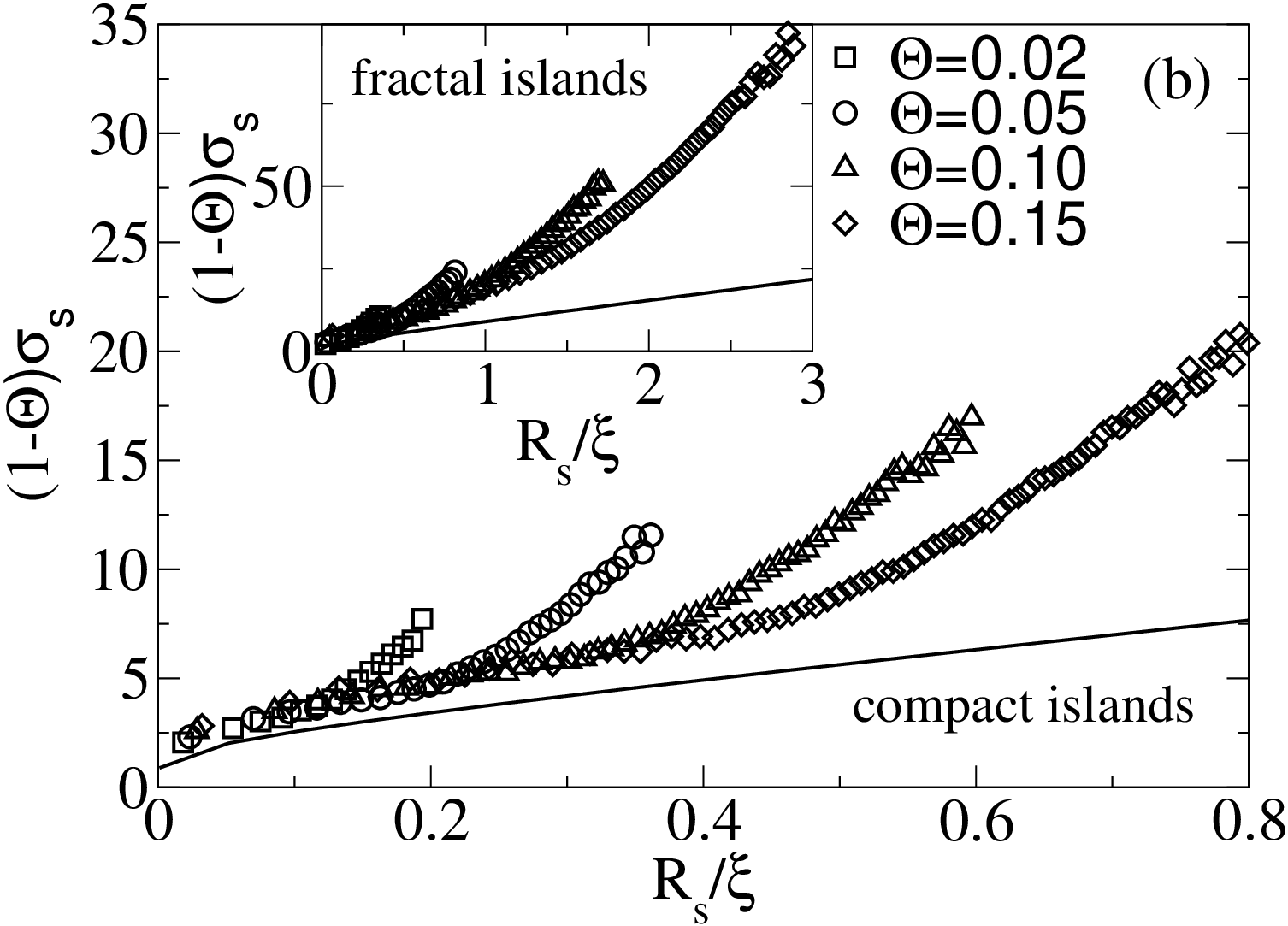}
\caption{Scaling plot of $(1-\Theta)\sigma_s$ as a function of
  $R_s/\xi$ for the model representing compact island morphologies,
  with both $R_s$ and $\xi^{-2}=2 \sigma_1 n_1 + \sum_{s\geq 2}
  \sigma_s n_s$ determined from the KMC simulations, (a) for various
  $\Gamma$ and fixed $\Theta=0.2$, and (b) for
  various $\Theta$ and fixed $\Gamma=10^8$.  The
  insets in (a) and (b) show the corresponding results for the model
  representing fractal island morphologies.  The solid lines represent
  the specific functional form in Eq.~(\ref{eq:sigma_SC}) predicted by
  the self-consistent theory.}
 \label{fig:sigma-rxi}
\end{figure}

\begin{figure*}[t!]
\includegraphics[width=0.32\textwidth]{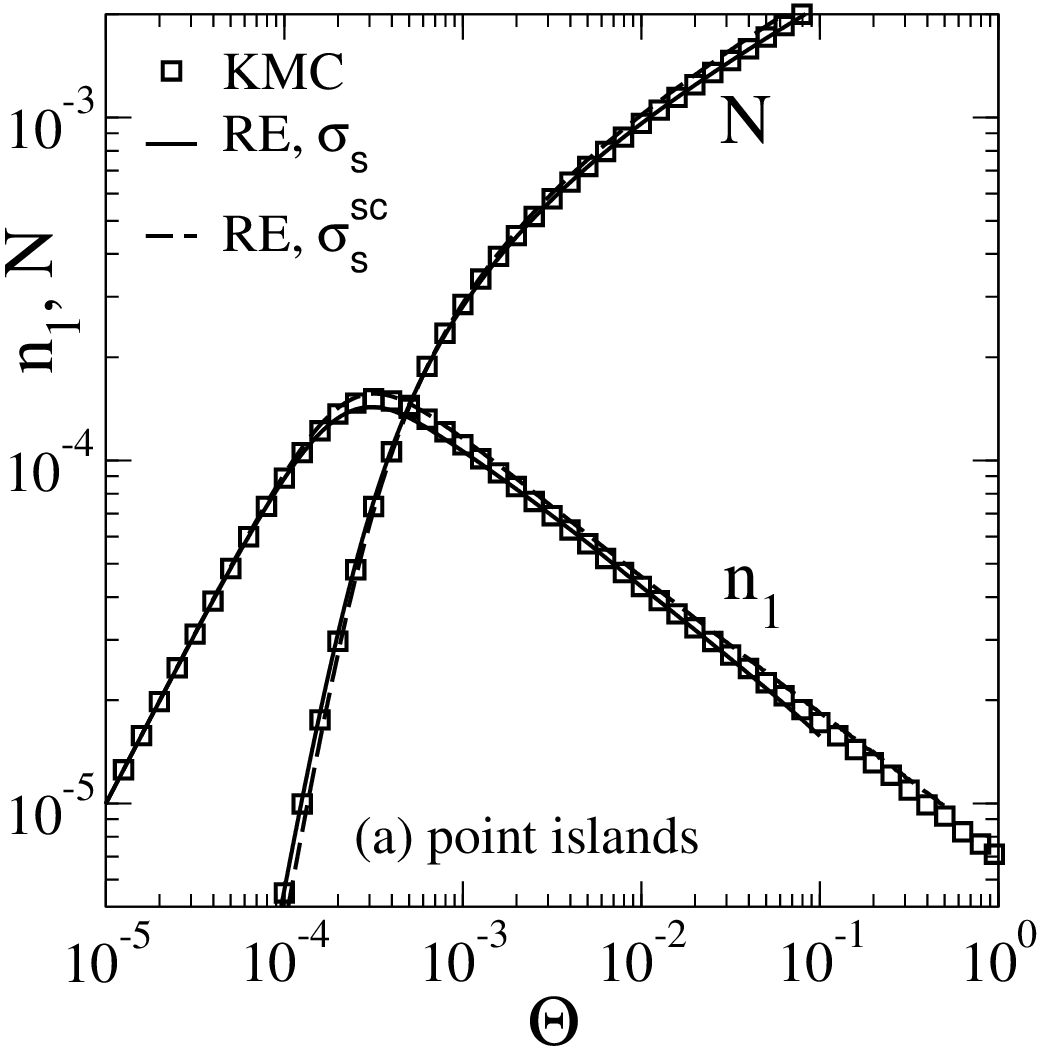}
\hspace*{0.01\textwidth}\includegraphics[width=0.32\textwidth]{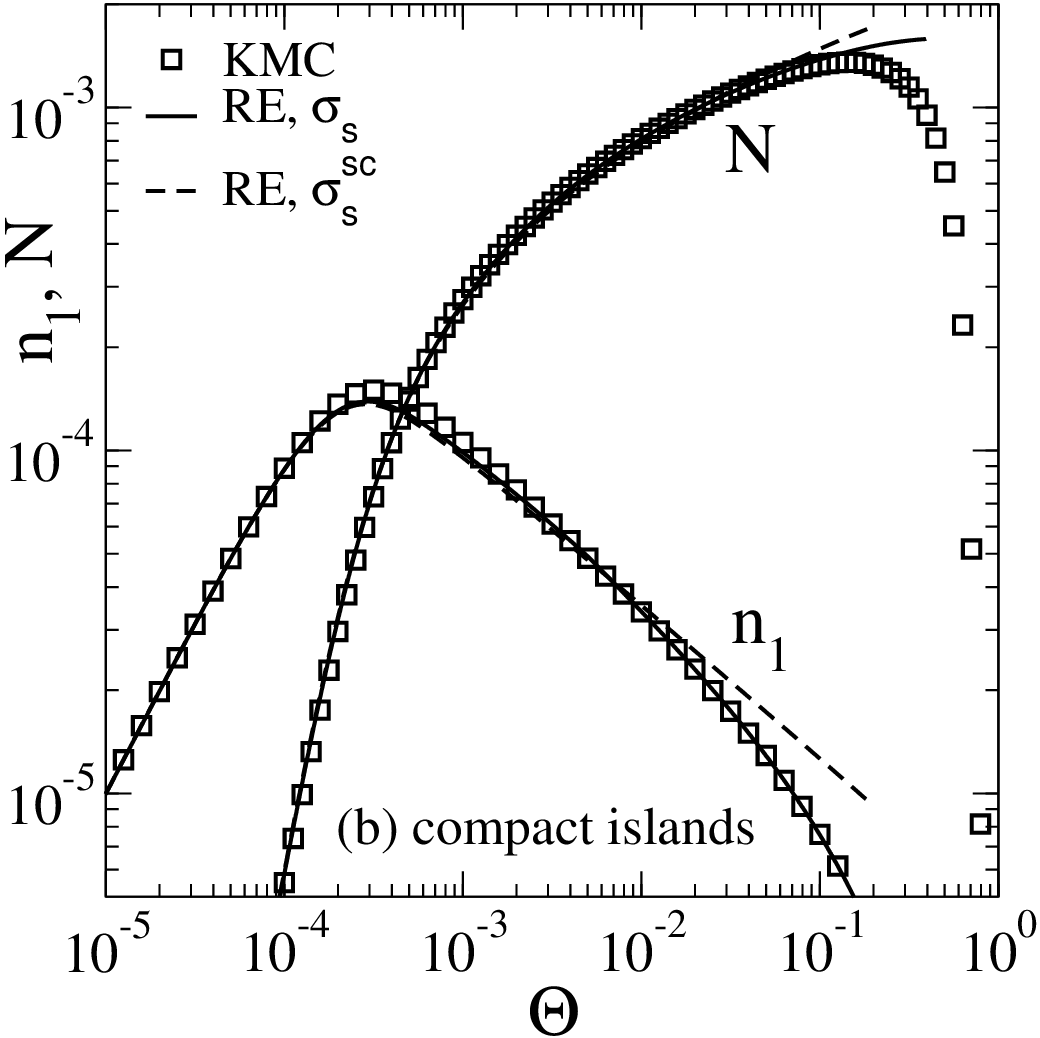}
\hspace*{0.01\textwidth}\includegraphics[width=0.32\textwidth]{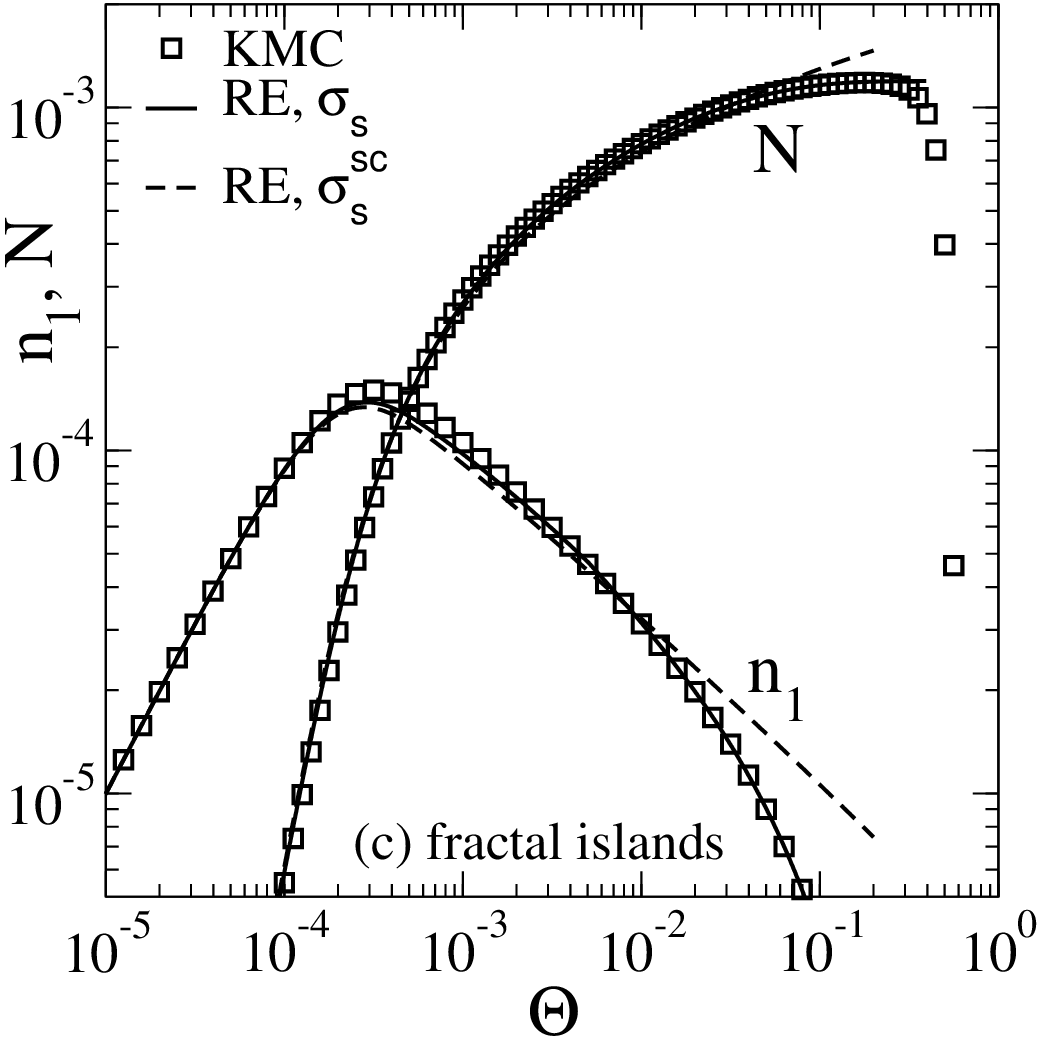}
\caption{KMC results (symbols) for the adatom density $n_1$ and island
  density $N$ as function of $\Theta$ at $\Gamma=10^7$ for (a) point,
  (b) compact, and (c) fractal island morphologies in comparison with
  the RE solutions, when using the direct capture areas $\kappa_s$ and
  capture numbers $\sigma_s$ from Sec.~\ref{sec:capture_numbers}
  (solid lines) and when using the capture numbers $\sigma^{\rm sc}_s$
  from the self-consistent theory according to Eq.~(\ref{eq:sigma_SC})
  (dashed lines).}
 \label{fig:ch4}
\end{figure*}

To determine the $\sigma_s^{\rm sc}$, the RE (\ref{eq:n1}) and
(\ref{eq:ns}) are numerically solved with initial conditions $n_s=0$
at time $t=0$ and a cutoff value $s_c$ so that $n_s$ can be safely
neglected for $s>s_c$. In each integration step the implicit
Eq.~(\ref{eq:sigma_SC}) is solved for the $\sigma_s^{\rm sc}$.  The
results become sensitive to the island morphology via the dependence
of $R_s$ on $s$ in this approach. For point islands we take $R_s=1$
corresponding to one lattice constant. For the compact and fractal
island morphologies, we determined the mean radius of gyration of
islands of size $s$, as shown in Fig.~\ref{fig:rs}. The straight lines
in the double-logarithmic representation give $R_s\sim 0.42 s^{1/2}$
(compact islands) and $R_s\sim 0.47 s^{0.57}$ (fractal morphologies)
for large $s$. To compare the $\sigma_s^{\rm sc}$ with the $\sigma_s$
obtained from the KMC simulations, we used the full dependence of the
$R_s$ on $s$, i.e.\ including the small $s$ behavior, in our
integration of the RE. The results from the self-consistent theory are
shown in Fig.~\ref{fig:ch3-sigma} (open symbols). As can be seen from
the figure, the $\sigma_s^{\rm sc}$ deviate strongly from the KMC
results, both in their size and in their functional form. In
particular the self-consistent theory underestimates the capture
numbers for large $s$, as known from earlier work in the literature
\cite{Evans/etal:2006}.

It is interesting to see, whether the scaling of
$(1-\Theta)\sigma_s(\Theta,\Gamma)$ with $R_s/\xi$ is valid, if the
$\sigma_s$ and $n_s$ from the KMC simulations are used in the
expression for $\xi^{-2}$ in Eq.~(\ref{eq:sigma_SCb}). In this case
the linearization step used in this theory for deriving a linear
diffusion equation for the local adatom density $n_1(r)$ could be
reasoned, i.e.\ the step, where the term $2\sigma_1 n_1(r)^2+
\sum_{s>1}\sigma_s n_1(r)n_s(r)$ is replaced by $n_1(r)\xi^{-2}$ with
$\xi^{-2}=2\sigma_1 n_1+\sum_{s>1}\sigma_s n_s$ given by the mean
($r$-independent) densities (see Ref.~\onlinecite{Bales/Chrzan:1994}
for details). In Fig.~\ref{fig:sigma-rxi}
$(1-\Theta)\sigma_s(\Theta,\Gamma)$ is plotted as function of
$R_s/\xi$ for the models representing compact and fractal island
morphologies.  Figure~\ref{fig:sigma-rxi}(a) shows that indeed a data
collapse is obtained for different $\Gamma$ values at fixed
$\Theta$. However, with respect to the $\Theta$-dependence, tested in
Fig.~\ref{fig:sigma-rxi}(b), no scaling behavior is found. This
indicates that the linearization step in the self-consistent theory
leads to the unsatisfactory capture numbers. It has been shown that
correlation effects between island sizes and capture areas need to be
taken into account to improve theories for capture numbers and island
size distributions. This can been achieved by considering the joint
probability of island size and capture
area.\cite{Amar/etal:2001,Popescu/etal:2001a,
  Mulheran/Robbie:2000,Evans/Bartelt:2002,Mulheran:2004}

\section{Adatom and island densities}
\label{sec:adatom_and_island_density}
Numerical integration of the RE with the $\kappa_s$ and $\sigma_s$
from Sec.~\ref{sec:capture_numbers} gives an excellent description of
the adatom density $n_1$ and of the island density $N$ as a function
of $\Theta$ and $\Gamma$ for all island morphologies in the
pre-coalescence regime. This is demonstrated in Fig.~\ref{fig:ch4},
where $n_1$ and $N$ from the KMC simulation (open squares) and RE
solution (solid lines) are plotted as function of $\Theta$ for
$\Gamma=10^7$. For compact and fractal island morphologies the KMC
data for $N$ steeply fall for coverages larger than 15\% (compact
islands) and 30\% (fractal islands) because of island coalescences.
Small deviations of the RE solution for $n_1$ can be seen close to
its maximum, where it slightly underestimates the adatom density.  The
agreement for the other simulated $\Gamma$ values is of the same
quality.  As known from previous studies,\cite{Bales/Chrzan:1994,
  Brune:1998, Popescu/etal:2001a, Ratsch/Venables:2003} the RE
predict $n_1$ and $N$ quite well also, when using the self-consistent
capture numbers from Eq.~(\ref{eq:sigma_SC}). The corresponding
solutions are drawn as dashed lines in Fig.~\ref{fig:ch4}. In view of
the discrepancies discussed in Sec.~\ref{sec:capture_numbers}, this
good predictive power of the RE under use of the self-consistent
capture numbers $\sigma^{\rm sc}_s$ is surprising.

\section{Island size distributions} 
\label{sec:ISD}
Since the $\sigma^{\rm sc}_s(\Theta,\Gamma)$ deviate strongly from the
$\sigma_s(\Theta,\Gamma)$, the RE with self-consistent capture
numbers fail to predict the ISD. This failure was reported already
when the self-consisting theory was developed.\cite{Bales/Chrzan:1994}
In the following we therefore do no longer consider the
self-consistent theory, but concentrate on the principal questions
whether the RE with the capture numbers $\sigma(\Theta,\Gamma)$ are
successful in predicting the ISD, and if so, whether in the limit
$\Gamma\to\infty$ the asymptotic form (\ref{eq:asympotic_form}) for
the scaling function becomes valid. In this section we address the
first of these two questions.

\begin{figure}[t!]
\includegraphics[scale=0.3]{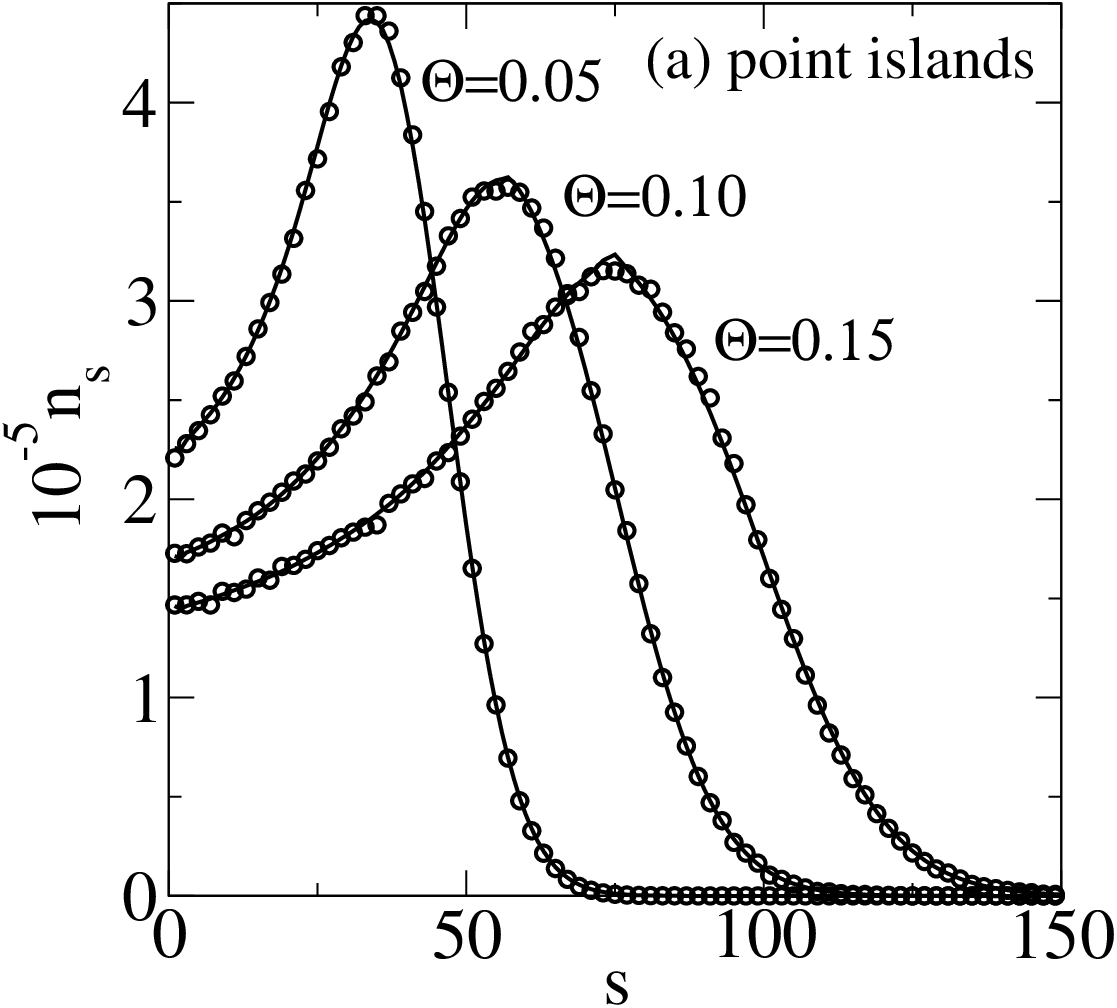}
\includegraphics[scale=0.3]{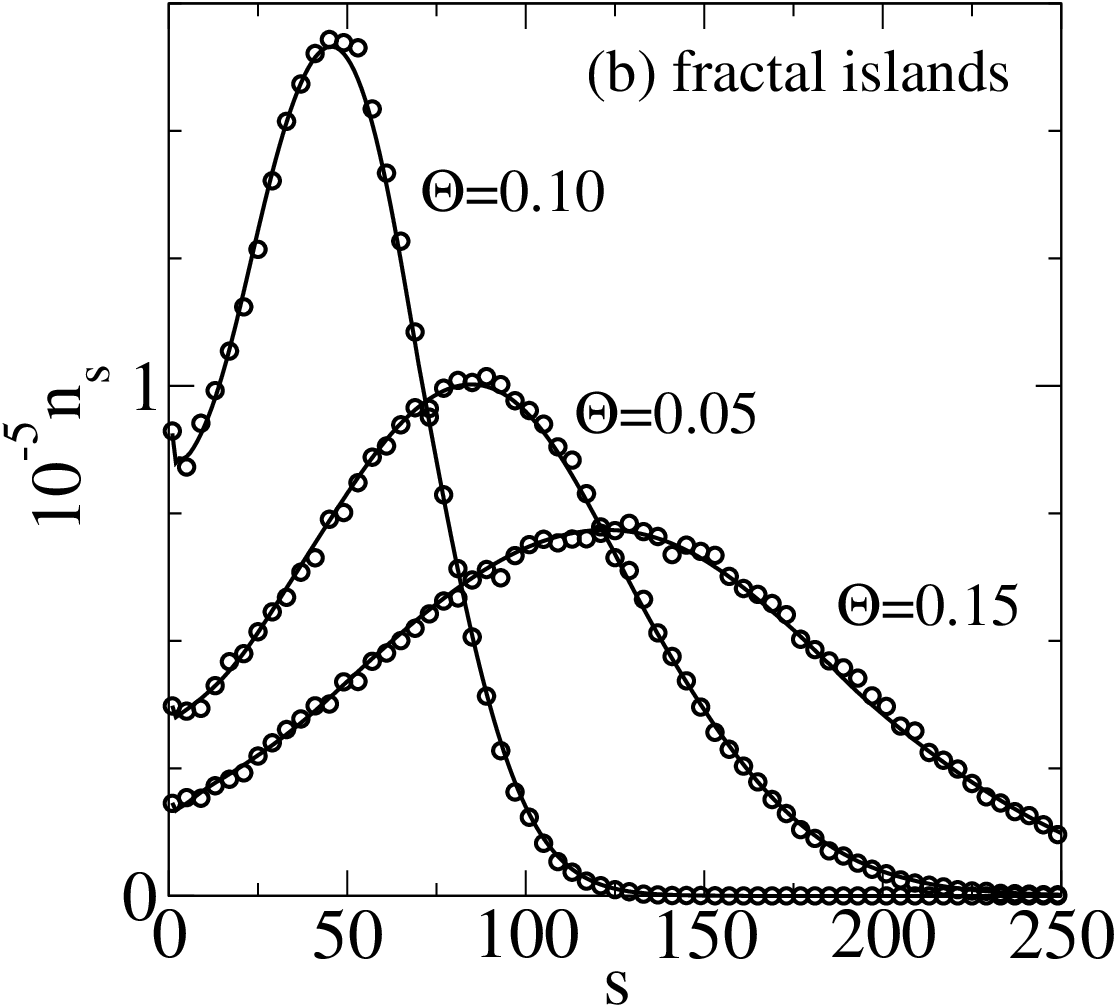}
\caption{Simulated island size distribution (circles) in comparison
  with the RE solution (lines) for three different $\Theta$ at
  $\Gamma=10^7$ for the models representing (a) point and (b) fractal
  island morphologies.}
 \label{fig:ch5_ISD}
\end{figure}

\begin{figure}[t!]
 \includegraphics[scale=0.3]{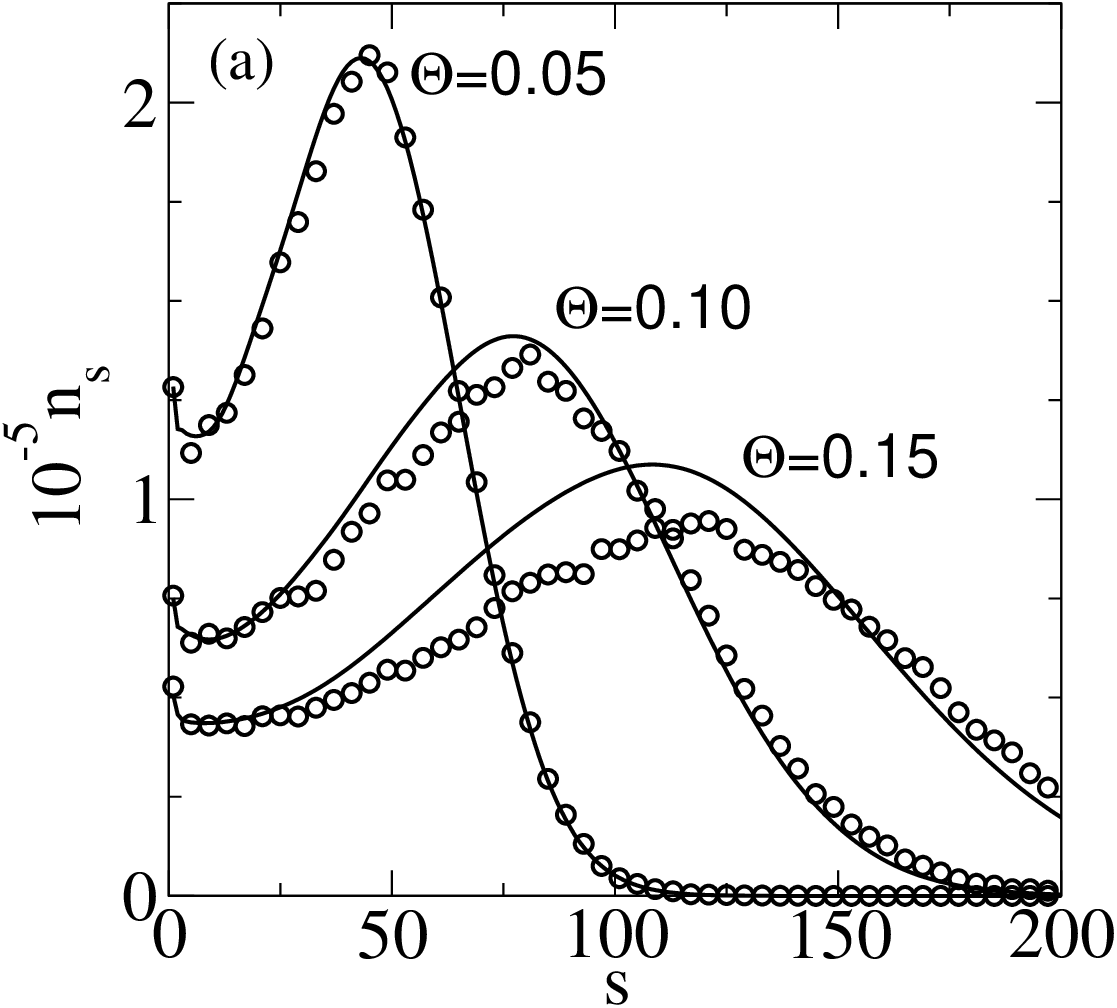}
 \includegraphics[scale=0.3]{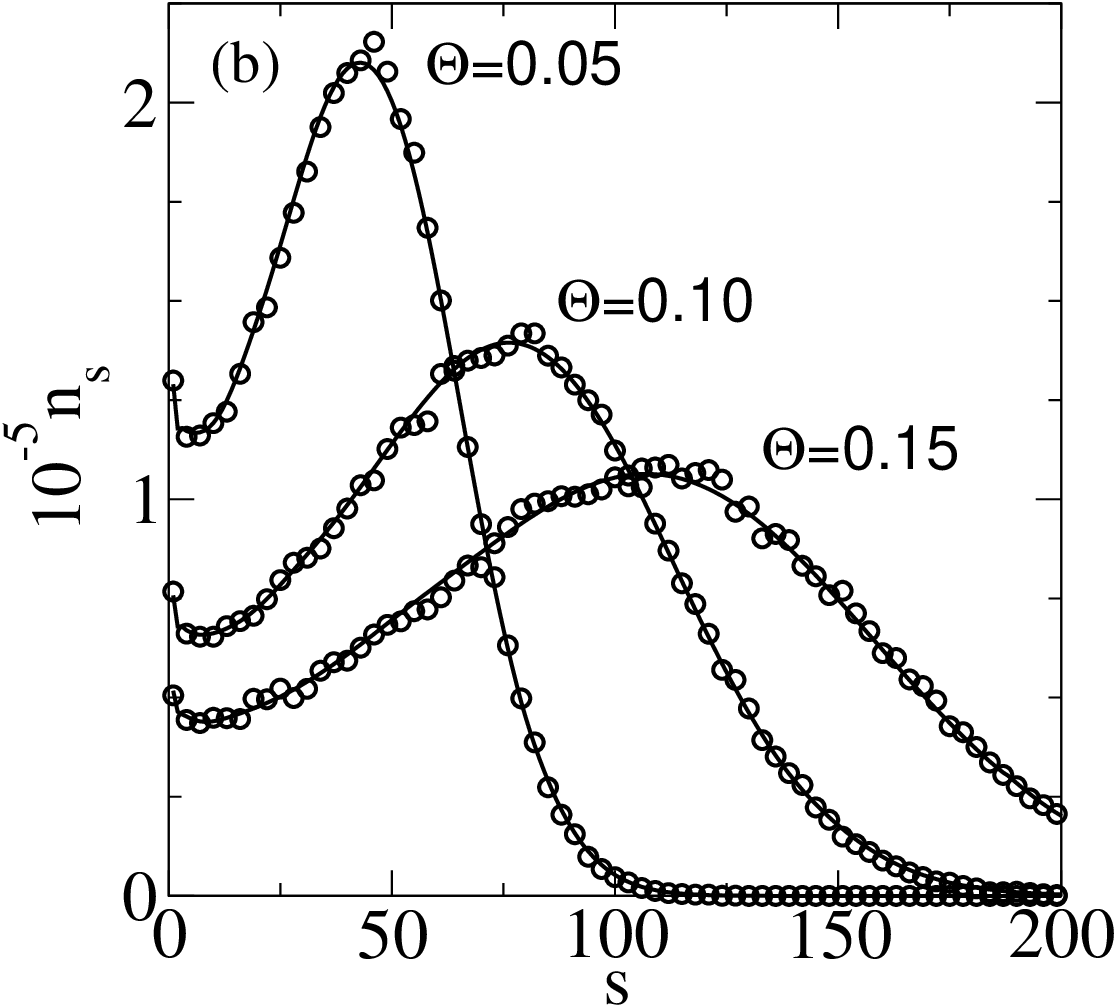}
 \caption{(a) ISD and (b) sub-ISD for compact islands obtained from
   the KMC simulations (circles) in comparison with the RE solution
   (lines) for $\Gamma=10^7$ and three different coverages $\Theta$.}
 \label{fig:ch5_ISD_compact}
\end{figure}

Representative results for the ISD (symbols) in comparison with the
RE predictions are shown in Fig.~\ref{fig:ch5_ISD} for $\Gamma=10^7$
and three different coverages $\Theta$, for point and fractal island
morphologies. The excellent agreement between the RE predictions and
the KMC data in that figure is also found for the other simulated $\Gamma$
values. As was shown in Ref.~\onlinecite{Koerner/etal:2010} for the
fractal island morphologies, a $\chi^2$ test with a standard
significance level of 5\% is passed up to a coverage of $\Theta=0.18$.
For larger $\Theta$, coalescence events, not included in the RE
approach, become relevant.

For the compact island morphologies, a good agreement of the KMC data
with the RE prediction is obtained up to coverages of about
$\Theta=0.05$ only, see Fig.~\ref{fig:ch5_ISD_compact}a. The reason
for the discrepancies are coalescence events that become important
already for small $\Theta\gtrsim0.05$, in contrast to what one may
conclude from the behavior of the mean island density shown in
Fig.~\ref{fig:ch4}, where coalescences seem to be irrelevant up to
coverages of about 15\%. One can take out the coalescence effect in
the calculation of the ISD by following the islands in the simulations
and by counting coalesced islands as if they were separated. The
islands identified in this way were referred to as sub-islands and the
resulting ISD as sub-ISD in Refs.~\onlinecite{Bartelt/Evans:1993,
  Evans/etal:2006}. In the same way as described in Sec.~\ref{sec:II}
we determined the $\kappa_s'$ and $\sigma_s'$ for the sub-islands and
integrated the RE (\ref{eq:n1}) and (\ref{eq:ns}) with these input
quantities. As shown in Fig.~\ref{fig:ch5_ISD_compact}(b), these RE
results for the sub-ISD give again excellent agreement with the KMC
data.

\begin{figure}[t!]
 \includegraphics[width=7cm]{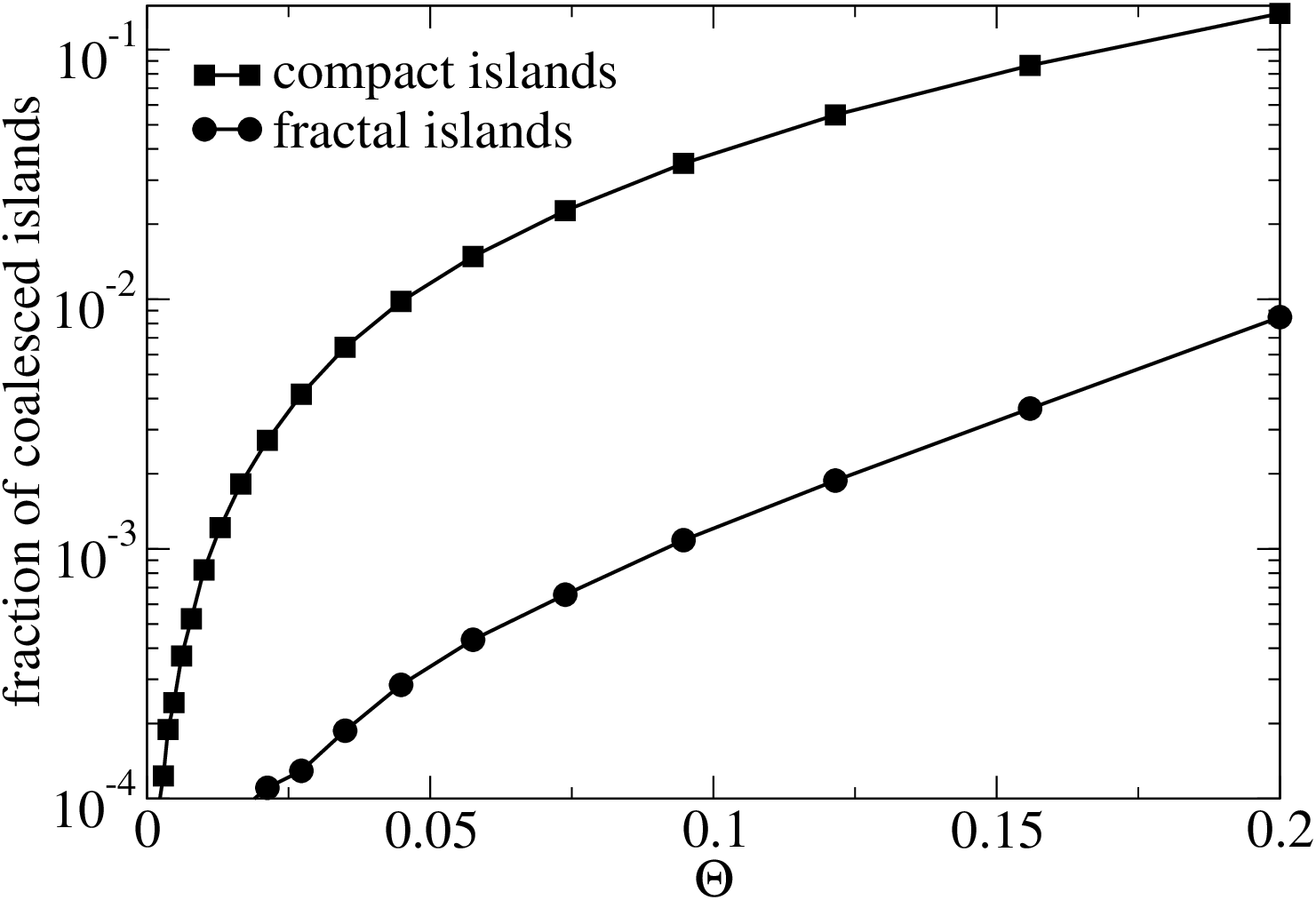}
 \caption{Fraction of coalesced islands as a function of the coverage
   $\Theta$ for $\Gamma=10^7$.}
\label{fig:fig6}
\end{figure}

That coalescence events are much more frequent for compact than for
fractal islands is shown in Fig.~\ref{fig:fig6}, where we plotted the
fraction of the coalesced islands as a function of $\Theta$ both for
the compact and fractal island morphologies. This fraction was
determined by dividing the total number of coalescences up to the
coverage $\Theta$ by the total number of islands at this $\Theta$
value, i.e.\ islands which have undergone more than one coalescence
are counted with their corresponding multiplicities. As can be seen
from Fig.~\ref{fig:fig6}, the fraction of coalesced islands for
compact islands has already at $\Theta=0.05$ reached a level
comparable to that found for the fractal islands at $\Theta=0.2$.

The reason for the less frequent coalescences of fractal islands is
that two approaching fractal islands can avoid each other for some
time, because fingers of one islands grow into breaches between
fingers of the other island. When a finger enters a breach, its
further growth slows down because of the shielding inside the
breach. This screening effect and its consequence for coalescences has
been discussed earlier in the
literature.\cite{Brune:1998,Brune/etal:1999} A quantitative analysis
of the coalescence behavior of compact and fractal island morphologies
is given in Appendix~\ref{app:B}.

\section{Limiting behavior for $\Gamma \rightarrow \infty$}
\label{sec:Scaling_theory}
Based on our first key finding that for all morphologies and for all
coverages in the pre-coalescence regime, the ISDs from the KMC
simulations are successfully predicted by the RE we now turn to the
question, whether the scaled ISDs approach the asymptotic form
(\ref{eq:asympotic_form}) in the limit $\Gamma\to\infty$.

To answer this question is not easy because of various subtleties,
which let us revisit the derivation of the scaling function by Bartelt
and Evans \cite{Bartelt/Evans:1996, Evans/Bartelt:2001} in
Appendix~\ref{app:A}. As mentioned in the Introduction,
Eq.~(\ref{eq:scaling_form}) for the limiting curve of the scaled ISD
$f_\infty(x)$ should be valid if $f_\infty(x,\Theta)$ from
Eq.~(\ref{eq:limiting_curve}) is independent of $\Theta$. This is the
case if $C_\infty(x,\Theta)=\lim_{\Gamma\to\infty}\sigma_{x\bar
  s}(\Theta,\Gamma)/\bar\sigma(\Theta,\Gamma)$ for the scaled capture
numbers and $z_\infty(\Theta)=\lim_{\Gamma\to\infty}\partial\ln\bar
s(\Theta,\Gamma)/\partial\ln\Theta$ also have $\Theta$-independent
limits.  A further requirement for the validity of
Eq.~(\ref{eq:asympotic_form}) is that
$\lim_{\Gamma\to\infty}\bar\kappa(\Theta,\Gamma)/\bar
s(\Theta,\Gamma)=0$, where $\bar\kappa=N^{-1} \sum_{s>1}\kappa_sn_s$
is the mean direct capture area. This condition can be expected to be
fulfilled for compact and point island morphologies and is in fact the
reason, why the scaling function of the direct capture areas should
not enter the RE prediction (\ref{eq:asympotic_form}). If
$\lim_{\Gamma\to\infty}\bar\kappa(\Theta,\Gamma)/\bar
s(\Theta,\Gamma)>0$, $f_\infty(x,\Theta)$ can be expected to depend on
$\Theta$ and one would need to solve the semi-linear partial
differential equation (\ref{eq:finf-pde}) for $f_\infty(x,\Theta)$.
Note that the $\kappa_s$ cannot increase stronger than linearly with
$s$, and accordingly $\bar\kappa$ should not increase more than
linearly with $\bar s$.

In interpreting numerical results for finite $\Gamma$, we have to pay
attention to the fact that for smaller $\Theta$ larger $\Gamma$ values
are needed to approach the limiting curves. This is because $\bar
s(\Theta,\Gamma)$ must become large enough to reach the ``continuum
limit'' (and larger $\Gamma$ are needed to obtain the same $\bar s$ at
smaller $\Theta$), and because the relation
$n_1\sim(1-\Theta)/\Gamma\bar\sigma N$, used in the derivation of
Eq.~(\ref{eq:asympotic_form}), should be obeyed. This relation is
usually referred to as the quasi-stationary condition, since it
follows from balancing the adatom attachment rate $D\bar\sigma nN$ to
islands with the deposition rate $F(1-\Theta)$. However, as was shown
earlier,\cite{Dieterich/etal:2008-2} the relation is also valid for
small $\Theta$ values in the regimes, where relative changes of $N$
are still large and have not leveled off. A refined scaling analysis
\cite{Einax/etal:2012} yields that, for $i=1$ as relevant here, the
relation holds for $(\Theta^2\Gamma)^{1/3}\gg1$,
implying again that for smaller $\Theta$ larger $\Gamma$ are needed to
identify the limiting behavior.

\begin{figure}[t!]
 \includegraphics[scale=0.3]{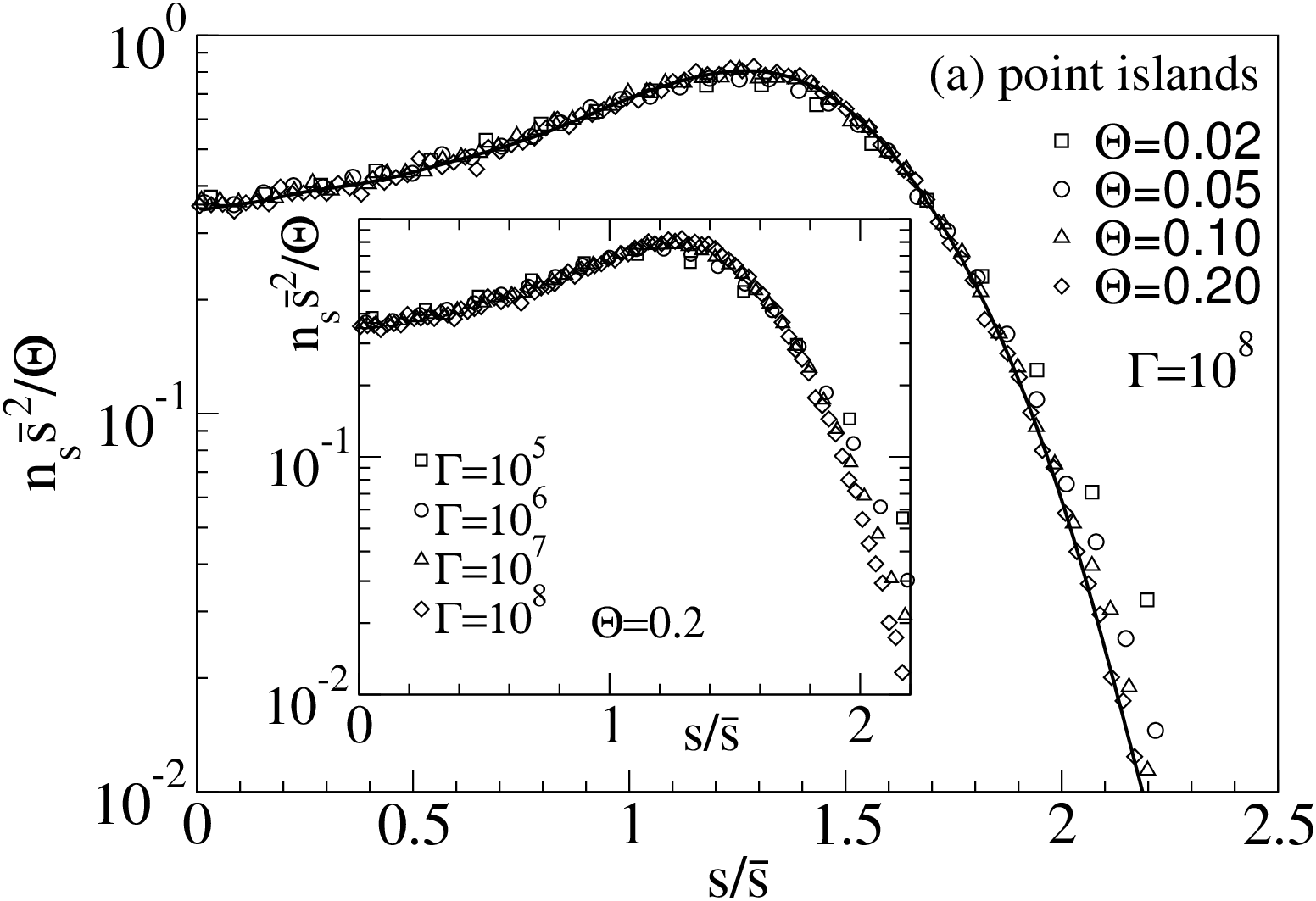}
 \includegraphics[scale=0.3]{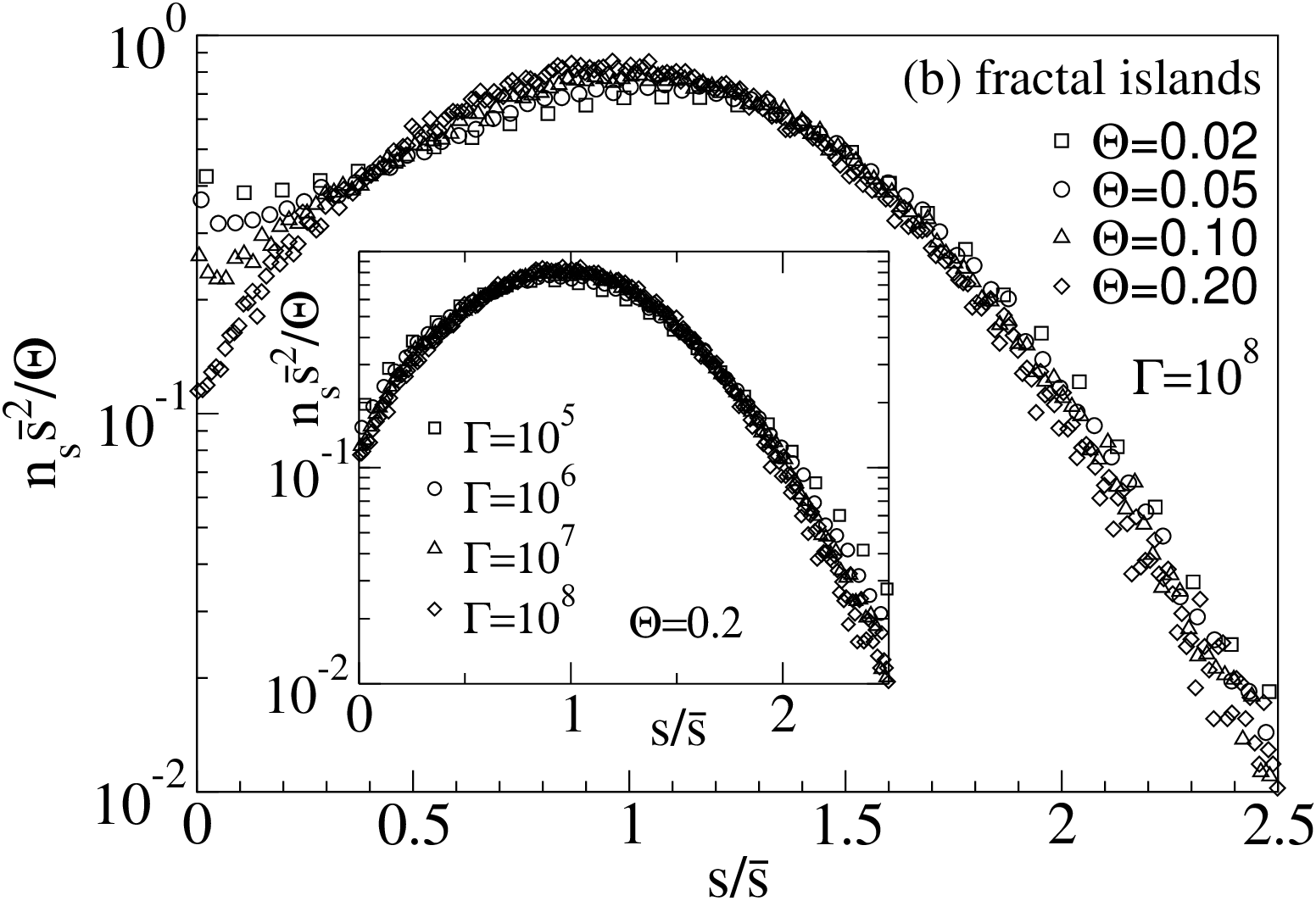}
 \caption{Scaled island size distributions $n_s\bar{s}^2/\Theta$ as
   function of the scaled island size $x=s/\bar s$ for fixed
   $\Gamma=10^8$ and four coverages $\Theta$, for (a) point and (b)
   fractal island morphologies. The insets show the scaled ISDs at
   fixed $\Theta=0.2$ and the four simulated $\Gamma$ values. The line
   in (a) is a fit to the data for $\Theta=0.2$ and $\Gamma=10^8$ and
   agrees with the analytical result Eq.~(\ref{eq:asympotic_form}),
   when using the line in Fig.~\ref{fig:scaled_sigma}(a) as the
   estimate for $C_\infty(x)$.}
 \label{fig:scaled_ISD}
\end{figure}

\begin{figure}[t!]
\includegraphics[scale=0.3]{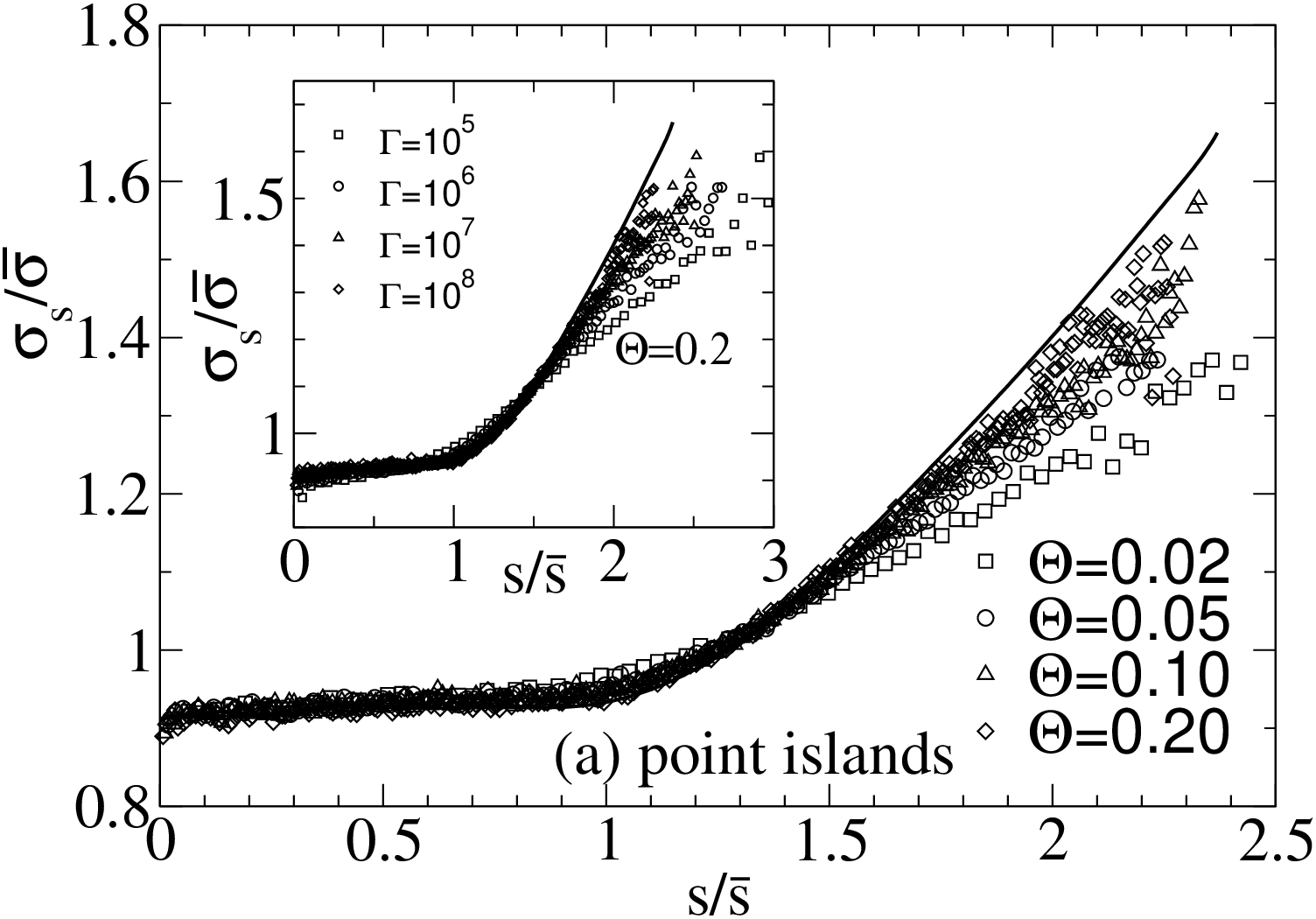}
\includegraphics[scale=0.3]{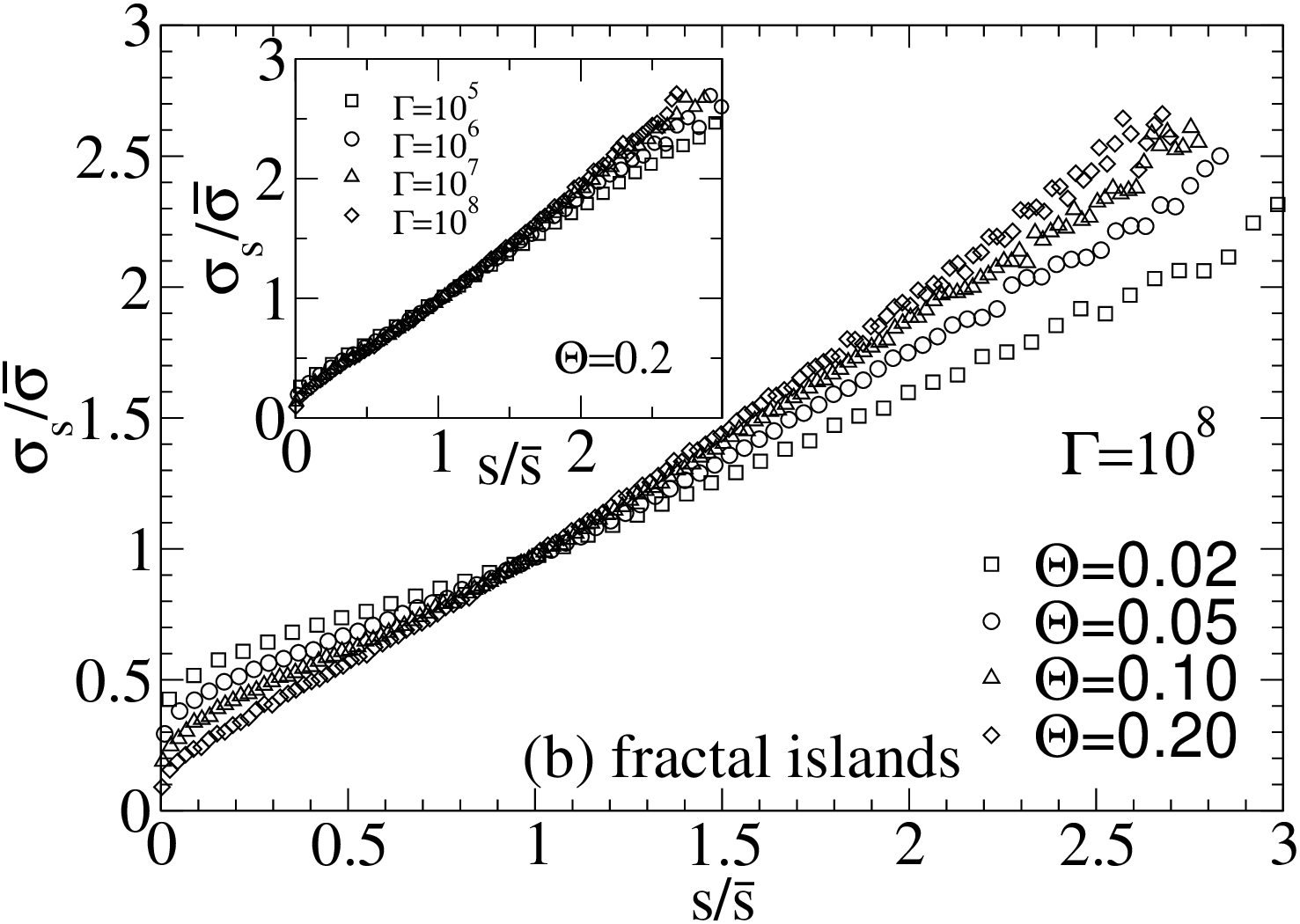}
\caption{Scaled capture numbers $\sigma_s/\bar\sigma$ as function of
  the scaled island size $s/\bar s$ for fixed $\Gamma=10^8$ and four
  coverages $\Theta$, for (a) point and (b) fractal island
  morphologies. The insets show the scaled capture numbers at fixed
  $\Theta=0.2$ and the four simulated $\Gamma$ values. The line in (a)
  marks the solution obtained from a numerical integration of
  Eq.~(\ref{eq:scaled_Capture_num}) when using the fit (line) to the
  scaled ISDs curve for $\Gamma=10^8$ and $\Theta=0.2$ in
  Fig.~\ref{fig:scaled_ISD}(a).}
 \label{fig:scaled_sigma}
\end{figure}

Figure~\ref{fig:scaled_ISD} shows $n_s\bar{s}^2/\Theta$ as a function
of $s/\bar{s}$ for $\Gamma=10^8$ at four different coverages for the
(a) point and (b) fractal island morphologies. In the insets the
scaled ISDs are shown for a fixed coverage $\Theta=0.2$ and different
$\Gamma$. In the case of the point island morphologies, the data
suggest the existence of a $\Theta$-independent limiting curve, in
agreement with previous findings.\cite{Evans/etal:2006} For the
fractal island morphologies, the scaled ISD for different $\Theta$
show no clear signature of a $\Theta$-independent limiting
curve. Based on the tendency of the simulated data for different
$\Theta$ and $\Gamma$ to become slightly closer to each other for
larger $\Theta$ and $\Gamma$, one may conjecture that also in this
case a limiting curve would be reached at larger $\Gamma$
values. However, the fact that for each fixed $\Theta$, the curves at
large $\Gamma$ are almost overlapping suggests that these are good
estimates of $f_\infty(x,\Theta)$. Our conclusion is therefore that it
is not likely that a $\Theta$-independent limiting curve exists for
the hit-and-stick model used here for the fractal island morphologies.

This conclusion is further corroborated by the fact that the scaled
direct capture areas exhibit a nearly linear dependence on $x$ for the
fractal islands (not shown). Thus we encounter the case here, where
the scaled direct capture areas $\bar\kappa(\Theta,\Gamma)/\bar
s(\Theta,\Gamma)$ appear to approach a non-vanishing limit for
$\Gamma\to\infty$, which would mean that in a strict treatment,
Eq.~(\ref{eq:asympotic_form}) can no longer be applied. If one
considers $f_\infty(x,\Theta)$ to depend only very weakly on $\Theta$,
$\partial f_\infty(x,\Theta)/\partial\Theta\simeq0$, we could replace
$C_\infty(x)$ by $C_{\rm tot}(x)=C_\infty(x)+\rho_\infty\Theta
K_\infty(x,\Theta)/(1-\Theta)$, where
$K_\infty(x,\Theta)=\lim_{\Gamma\to\infty}\kappa_{x\bar
  s}(\Theta,\Gamma)/\bar\kappa(\Theta,\Gamma)$ is the limiting curve
for the scaled direct capture areas and
$\rho_\infty=\lim_{\Gamma\to\infty}\bar\kappa(\Theta,\Gamma)/\bar
s(\Theta,\Gamma)>0$.

Our conclusions drawn with respect to the scaled ISDs of the point
islands are consistent with the behavior of the scaled capture
numbers, which are shown in Fig.~\ref{fig:scaled_sigma}(a) for the
same $\Theta$ and $\Gamma$ values as in
Fig.~\ref{fig:scaled_ISD}(a). In this Fig.~\ref{fig:scaled_sigma}(a)
an approach to a $\Theta$-independent limiting curve $C_\infty(x)$ can
be seen. In the case of the fractal island morphologies by contrast,
an approach to a $\Theta$-independent limiting curve cannot be clearly
identified, which gives further evidence that the $f_\infty(x,\Theta)$
are dependent on $\Theta$.

In order to test the validity of Eq.~(\ref{eq:asympotic_form}) for the
point islands, we set $z_\infty=2/3$ (see
Sec.~\ref{sec:capture_numbers}) and used a fit to the scaled ISD for
$\Gamma=10^8$ and $\Theta=0.2$ in Fig.~\ref{fig:scaled_ISD}(a) as an
estimate for $f_\infty(x)$.  The fit, which fulfills the constraints
of normalization and normalized first moment, is shown as line in this
figure.  We then estimated $C_\infty(x)$ based on this fit by
rewriting Eq.~(\ref{eq:finf-pde}) from Appendix~\ref{app:A} (for
$\Theta$-independent $f_\infty(x)$) in the form
\begin{align}
\label{eq:scaled_Capture_num}
\frac{\mathrm{d} C_\infty(x)}{\mathrm{d} x} &= (2z_\infty-1) -
[C_\infty(x)-z_\infty x] \frac{{\mathrm{d}} \ln
  (f_\infty(x))}{\mathrm{d}x} \, .
\end{align}
Since the solution of this differential equation is proportional to
$1/f_\infty(x)$, we preferred to integrate
Eq.~(\ref{eq:scaled_Capture_num}) with the initial condition
$C_\infty(0)=f_\infty(0)/(1-z_\infty)$ to achieve a stable numerical
results for large $x$ also.  The resulting estimate for
$C_{\infty}(x)$ is shown as line in
Fig.~\ref{fig:scaled_sigma}(a). The line lies slightly above the data
for the scaled capture numbers for $\Gamma=10^8$ and $\Theta=0.2$,
indicating that indeed an estimate of a limiting curve for the scaled
capture numbers is obtained. In a cross-check, we performed the
integral in Eq.~(\ref{eq:asympotic_form}) with the estimated
$C_{\infty}(x)$ and recovered the line in
Fig.~\ref{fig:scaled_ISD}(a).

\begin{figure}[ht!]
 \includegraphics[scale=0.3]{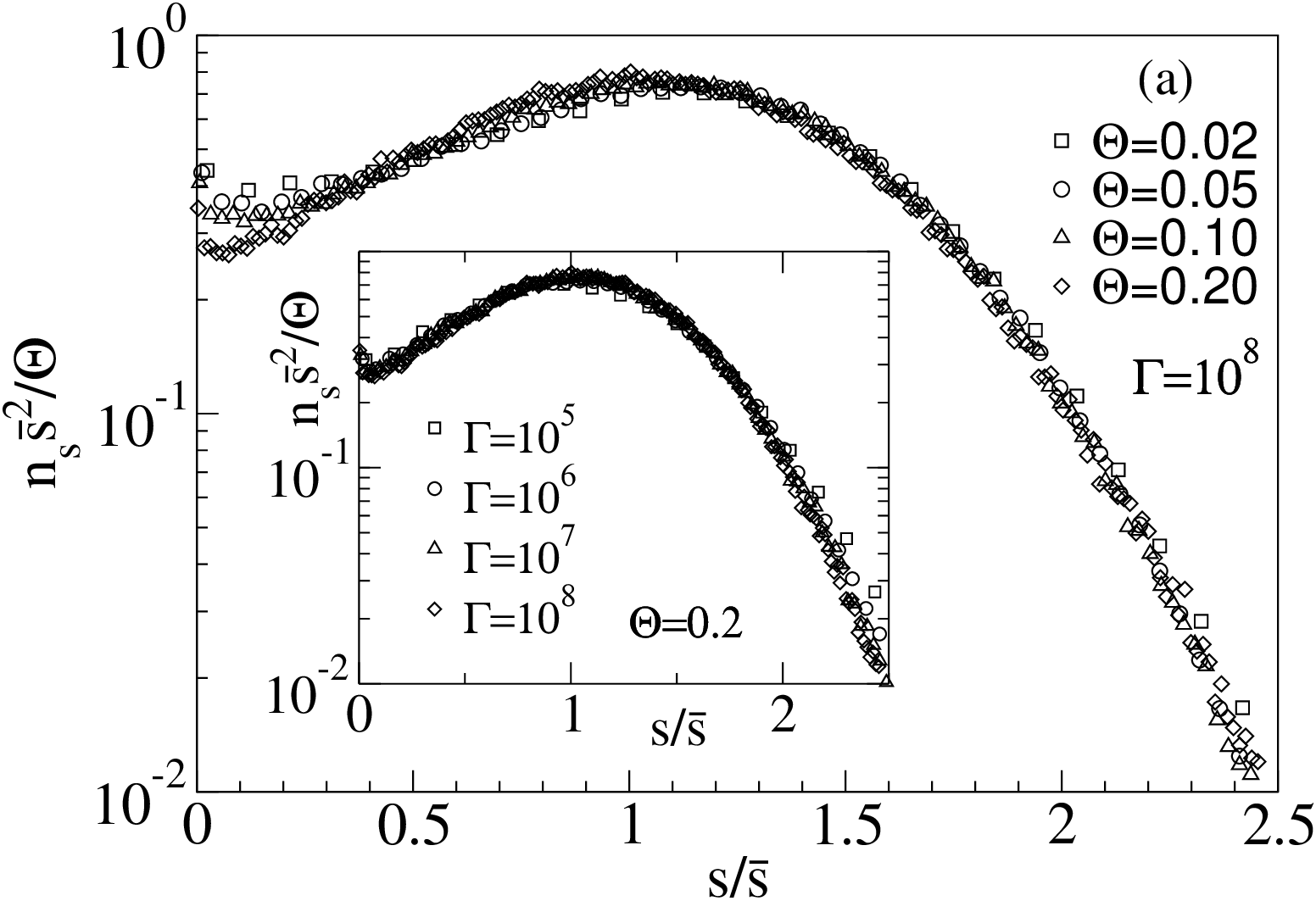}
 \includegraphics[scale=0.3]{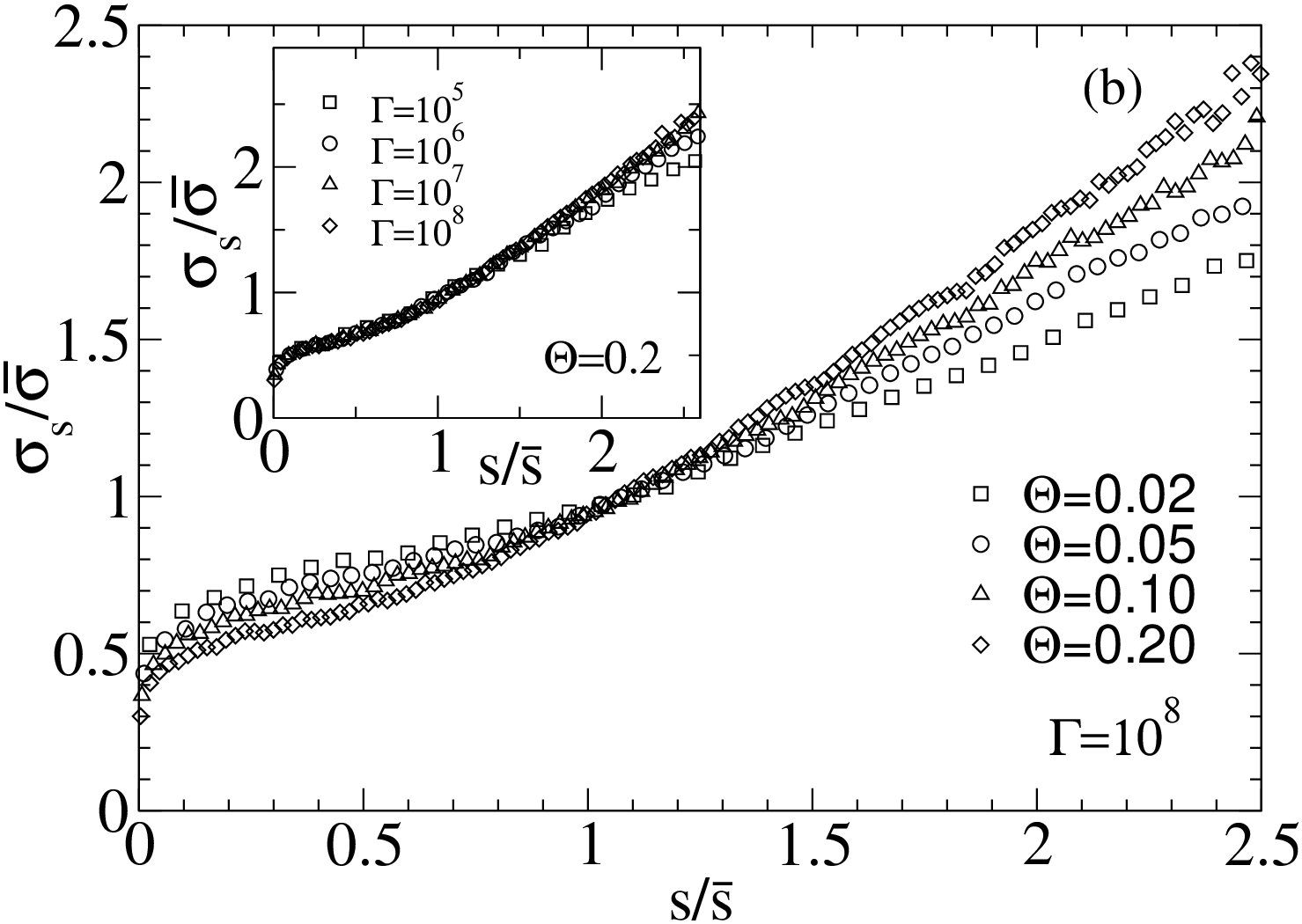}
 \caption{(a) Scaled island size distributions and (b) scaled capture
   numbers of sub-islands as function of the scaled island size in the
   case of compact island morphologies for fixed $\Gamma=10^8$ and
   four coverages $\Theta$. The inset shows the corresponding data at
   fixed $\Theta=0.2$ and the four simulated $\Gamma$ values.}
 \label{fig:scaled_subisd}
\end{figure}

For the compact island morphologies, Eq.~(\ref{eq:asympotic_form})
would be of limited practical use, because, as discussed in
Sec.~\ref{sec:ISD}, the RE (\ref{eq:n1}), (\ref{eq:ns}) fail to
predict the ISD correctly already at small $\Theta$ due to
coalescences. Nevertheless, from a conceptual viewpoint, it is
interesting to study the scaled ISD and their relation to the scaled
capture numbers for the sub-islands. The corresponding data shown in
Fig.~\ref{fig:scaled_subisd} indicate a behavior similar as for the
fractal island morphologies, where the limiting curves are dependent
on $\Theta$.

\section{Summary}
\label{sec:summary}
The capture numbers entering the RE for the growth kinetics of thin
films have been determined by KMC simulations in their dependence on
both the coverage $\Theta$ and the $\Gamma=D/F$ ratio for the point
island model and for two simple growth models representative for
islands with compact and fractal shapes. It was shown that the
$\Theta$-dependence of the capture numbers could not be accounted for
by the ratio $R_s/\xi$ of the mean island radius $R_s$ and the
effective adatom capture length $\xi$ of the RE. This suggests that
the strong deviations between the capture numbers determined from the
simulations and the ones predicted by the self-consistent theory have
their origin in the linearization step used in this theory. The RE
with self-consistent capture numbers nevertheless provide a good
quantitative account of the adatom and island density. The deviations
to the correct capture numbers lead, however, to a failure for a
description of the ISD.

Integration of the RE with the simulated capture numbers determined
from the KMC simulations gives an excellent quantitative prediction of
the ISDs.  For the compact islands morphologies, it was found that
coalescence events, not considered in the RE, become relevant
already at small coverages well below $\Theta\simeq0.15$, where
coalescence events do not significantly affect the island
density. Compared to the fractal island morphologies, the coalescence
rate for the compact morphologies is much higher. The ISD is affected
already by a rather small number of coalescences, because these lead
to a reshuffling of weights for different island sizes. The lower
coalescence rate for fractal morphologies is caused by the fact that
fingers of two approaching fractal islands typically first avoid each
other, which subsequently leads to a screening effect and a slowing
down of further growth of these fingers.

Finally we discussed the limiting curves for the scaled ISDs when
$\Gamma\to\infty$.  For the point islands the KMC data provide
evidence that these limiting curves are independent of the coverage,
which is given by the RE prediction (\ref{eq:asympotic_form}). This
means that there exists a true scaling behavior in the
$\Gamma\to\infty$ limit, where the dependence on $\Theta$ is fully
accounted for by the mean island size $\bar s$. For the growth models
representing compact and fractal island morphologies, the results
indicate that the limiting curves are dependent on $\Theta$.
This implies that one needs to solve the partial differential
equation (\ref{eq:finf-pde}) [or Eq.~(\ref{eq:finf-pde-2})]
to calculate $f_\infty(x,\Theta)$ from $C_\infty(x,\Theta)$.
Unfortunately no successful theory exists so far to predict the
limiting curve $C_\infty(x,\Theta)$ for the scaled capture numbers.

The limiting curves are also different for different
morphologies. Considering how sensitive the shape of the limiting
curves depends on the nonlinear behavior of the scaled capture numbers
as a function of the scaled island size, it is well possible that the
shape will also vary with details of the growth mechanisms, even if
the type of island morphology remains essentially the same.

\appendix

\section{Rate equation prediction of the limiting curves
for the scaled island size distribution}
\label{app:A}
For large $\Gamma$, $\bar s\sim N^{-1}\sim\Gamma^{1/3}$ and $x=s/\bar
s$ becomes a continuous variable, which allows one to derive a
determining equation for the scaled ISD in dependence of the scaled
capture numbers. The derivation was first presented by Bartelt and
Evans.\cite{Bartelt/Evans:1996, Evans/Bartelt:2001} Replacing the
variable $s$ by $x$ and using
$\partial/\partial(Ft)=(1-\Theta)\partial/\partial\Theta$,
Eqs.~(\ref{eq:ns}) can be written in the continuum limit as
\begin{align}
  \label{eq:ns_continuum}
\frac{\partial n_{x\bar s}}{\partial\Theta} &=-\frac{1}{(1-\Theta)\bar s}
\left\{\Gamma n_1
\frac{\partial}{\partial x} \Big(\sigma_{x\bar s} n_{x\bar s}\Big)
    +\frac{\partial}{\partial x}
      \Big(\kappa_{x\bar s} n_{x\bar s}\Big)\right\} \, .
\end{align}
Defining $f(x,\Theta,\Gamma)=\bar s^2n_{x\bar s}/\Theta$,
$C(x,\Theta,\Gamma)=\sigma_{x\bar s}/\bar\sigma$, and
$K(x,\Theta,\Gamma)=\kappa_{x\bar s}/\bar\kappa$, one has
\begin{align}
\frac{\partial}{\partial x} \Big(\sigma_{x\bar s} n_{x\bar s}\Big)&=
\frac{\Theta\bar\sigma}{\bar s^2}\frac{\partial(Cf)}{\partial x}
\label{eq:dxsigma}\\
\frac{\partial}{\partial x} \Big(\kappa_{x\bar s} n_{x\bar s}\Big)&=
\frac{\Theta\bar\kappa}{\bar s^2}\frac{\partial(Kf)}{\partial x}
\label{eq:dxkappa}\\
\frac{\partial n_{x\bar s}}{\partial\Theta} &=-(2z-1)f-
zx\frac{\partial f}{\partial x}+\Theta\frac{\partial f}{\partial\Theta}
\label{eq:dtn}
\end{align}
where $z(\Theta,\Gamma)=\partial\ln\bar s/\partial\ln\Theta$.  The
reduced RE moreover predict $n_1\sim(1-\Theta)/(\Gamma\bar\sigma
N)\sim(1-\Theta)\bar s/(\Theta\Gamma\bar\sigma)$ for large $\Gamma$
and fixed $\Theta>\Theta_x\sim\Gamma^{-1/2}$.  Inserting this relation
and Eqs.~(\ref{eq:dxsigma}, \ref{eq:dxkappa}, \ref{eq:dtn}) into
Eq.~(\ref{eq:ns_continuum}) gives
\begin{align}
(2z-1)f+zx\frac{\partial f}{\partial x}-
\Theta\frac{\partial f}{\partial\Theta}&=
\frac{\partial (Cf)}{\partial x}+
\frac{\Theta\bar\kappa}{(1-\Theta)\bar s}
\frac{\partial (Kf)}{\partial x}
\label{eq:f-pde}
\end{align}
Introducing the limits
$C_\infty(x,\Theta)=\lim_{\Gamma\to\infty}C(x,\Theta,\Gamma)$,
$K_\infty(x,\Theta)=\lim_{\Gamma\to\infty}K(x,\Theta,\Gamma)$ and
$z_\infty(\Theta)=\lim_{\Gamma\to\infty}z(\Theta,\Gamma)$,
Eq.~(\ref{eq:f-pde}) yields a determining equation for
$f_\infty(x,\Theta)=\lim_{\Gamma\to\infty}f(x,\Theta,\Gamma)$.

For $\lim_{\Gamma\to\infty}\bar\kappa/\bar s=0$ one obtains
\begin{align}
(2z_\infty-1)f_\infty+z_\infty x\frac{\partial f_\infty}{\partial x}-
\Theta\frac{\partial f_\infty}{\partial\Theta}&=
\frac{\partial (C_\infty f_\infty)}{\partial x}
\label{eq:finf-pde}
\end{align}
The condition $\lim_{\Gamma\to\infty}\bar\kappa/\bar s=0$ is valid for
point islands, and it can be expected to hold also for compact island
morphologies unless atoms deposited on top of islands are essentially
all attaching to the island edge in the first layer (a situation
unlikely due to second layer nucleation on larger islands).

When integrating Eq.~(\ref{eq:finf-pde}) over $x$ from zero to
infinity, the first, second and third term on the left hand side yield
$(2z_\infty-1)$, $z_\infty$ (after a partial integration) and zero,
respectively, because of the normalization of $f_\infty$. The right
hand side becomes $[-C_\infty(0,\Theta)f_\infty(0,\Theta)]$ (note that
for large $x$, $C_\infty\sim x$ and $f_\infty$ must decrease faster
than $x$ to be normalizable -- simulation results show that $f_\infty$
should in fact decay much faster). Accordingly, the relation
\begin{align}
f_\infty(0,\Theta)=\frac{1-z_\infty(\Theta)}{C_\infty(0,\Theta)}
\label{eq:f0-C0}
\end{align}
must be fulfilled. A corresponding relation can be derived in the same
way already from Eq.~(\ref{eq:f-pde}). Analogously, when first
multiplying Eq.~(\ref{eq:finf-pde}) with $x$ and then integrating,
one obtains
\begin{align}
\int_0^\infty C_\infty(x,\Theta)f_\infty(x,\Theta)=1 \, .
\label{eq:intfc}
\end{align}
Integrating Eq.~(\ref{eq:finf-pde}) to a finite value $x$ then
yields
\begin{align}
C_\infty(x,\Theta)&=z_\infty(\Theta)x+\frac{1-z_\infty(\Theta)}{f_\infty(x,\Theta)}
\int_x^\infty dx'\,f_\infty(x',\Theta)\nonumber\\
&{}-\frac{\Theta}{f_\infty(x,\Theta)}
\frac{\partial}{\partial\Theta}\int_0^x dx'\,f_\infty(x',\Theta)\,,
\label{eq:c-f}
\end{align}
which expresses $C_\infty(x,\Theta)$ as a functional of
$f_\infty(x,\Theta)$.

When one further assumes that the limiting curve $f_\infty$ is
independent of $\Theta$, one has $\partial f_\infty/\partial\Theta=0$
and can neglect the corresponding term in Eq.~(\ref{eq:f-pde}). For
self-consistency, this requires also $C_\infty$ and $z_\infty$ to
become independent of $\Theta$. In fact, one can conversely show that
if $C_\infty$ and $z_\infty$ are independent of $\Theta$, $f_\infty$
must by independent of $\Theta$ also. Under this assumption
Eq.~(\ref{eq:finf-pde}) then reduces to a separable ordinary
differential equation, whose solution is given by
Eq.~(\ref{eq:asympotic_form}), with $C_{\rm tot}(x)$ equal to
$C_\infty(x)$ and $z$ equal to $z_\infty$.

If there exists a finite limit $\rho_\infty(\Theta)=
\lim_{\Gamma\to\infty}\bar\kappa/\bar s>0$ , as it may be the case for
fractal island morphologies (see the discussion in
Sec.~\ref{sec:Scaling_theory}), Eq.~(\ref{eq:f-pde}) yields
\begin{align}
\label{eq:finf-pde-2}
(2z_\infty-1)f_\infty&+z_\infty x\frac{\partial f_\infty}{\partial x}-
\Theta\frac{\partial f_\infty}{\partial\Theta}=\\
&\frac{\partial (C_\infty f_\infty)}{\partial x}+
\rho_\infty\frac{\Theta}{(1-\Theta)}
\frac{\partial (K_\infty f_\infty)}{\partial x}\nonumber
\end{align}
as determining equation for $f_\infty(x,\Theta)$. Strictly speaking, a
$\Theta$-independent $f_\infty(x)$ should not exist then and one needs
to solve the semi-linear partial differential equation
(\ref{eq:finf-pde-2}). If one nevertheless makes the approximation
$\partial f_\infty/\partial\Theta\simeq0$ in Eq.~(\ref{eq:finf-pde-2})
and considers $C_\infty$ and $z_\infty$ to be independent of (or only
weakly dependent on) $\Theta$, one would obtain the weakly
$\Theta$-dependent solution Eq.~(\ref{eq:asympotic_form}) with $C_{\rm
  tot}=C_\infty+\rho_\infty\Theta K_\infty/(1-\Theta)$.

\begin{figure}[b!]
 \includegraphics[scale=0.23]{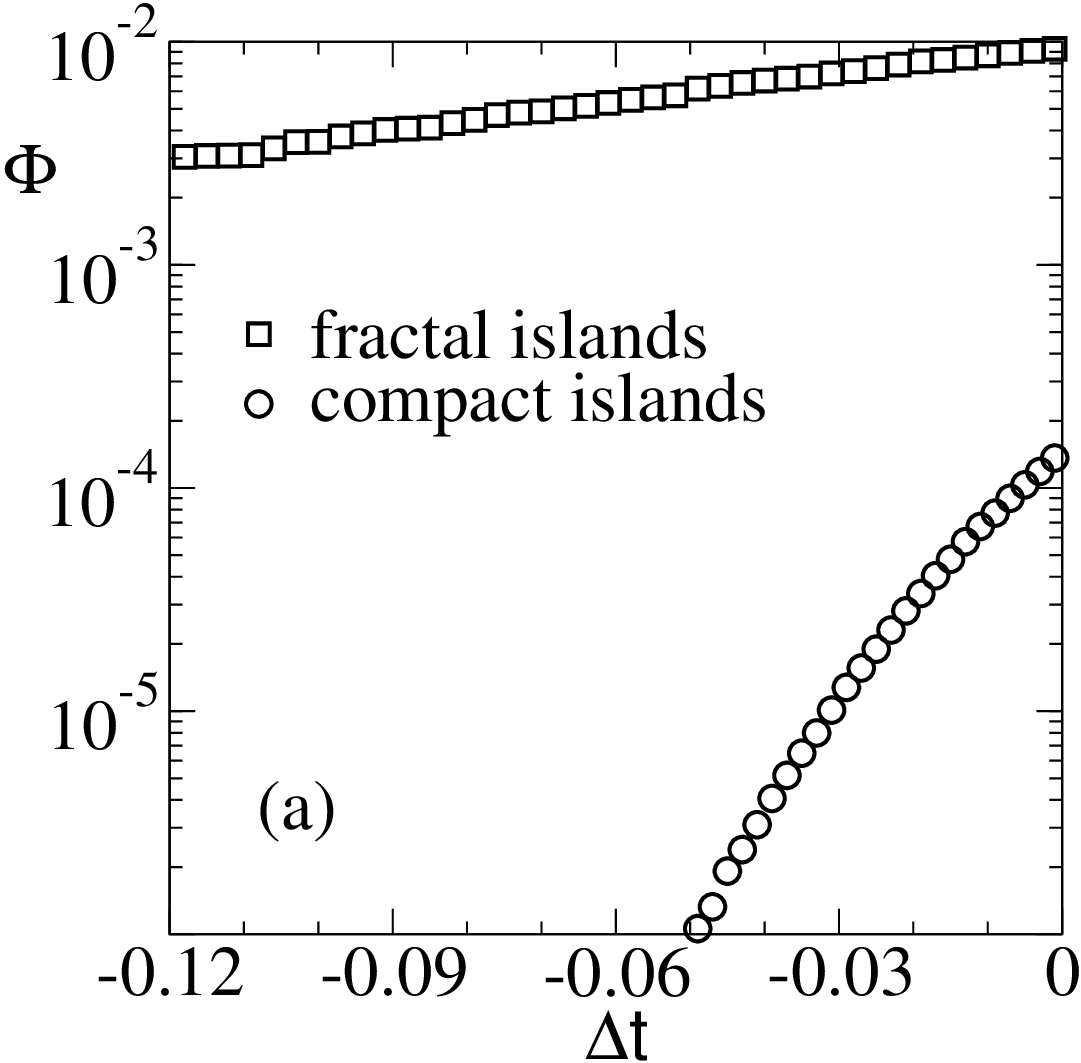}
 \includegraphics[scale=0.23]{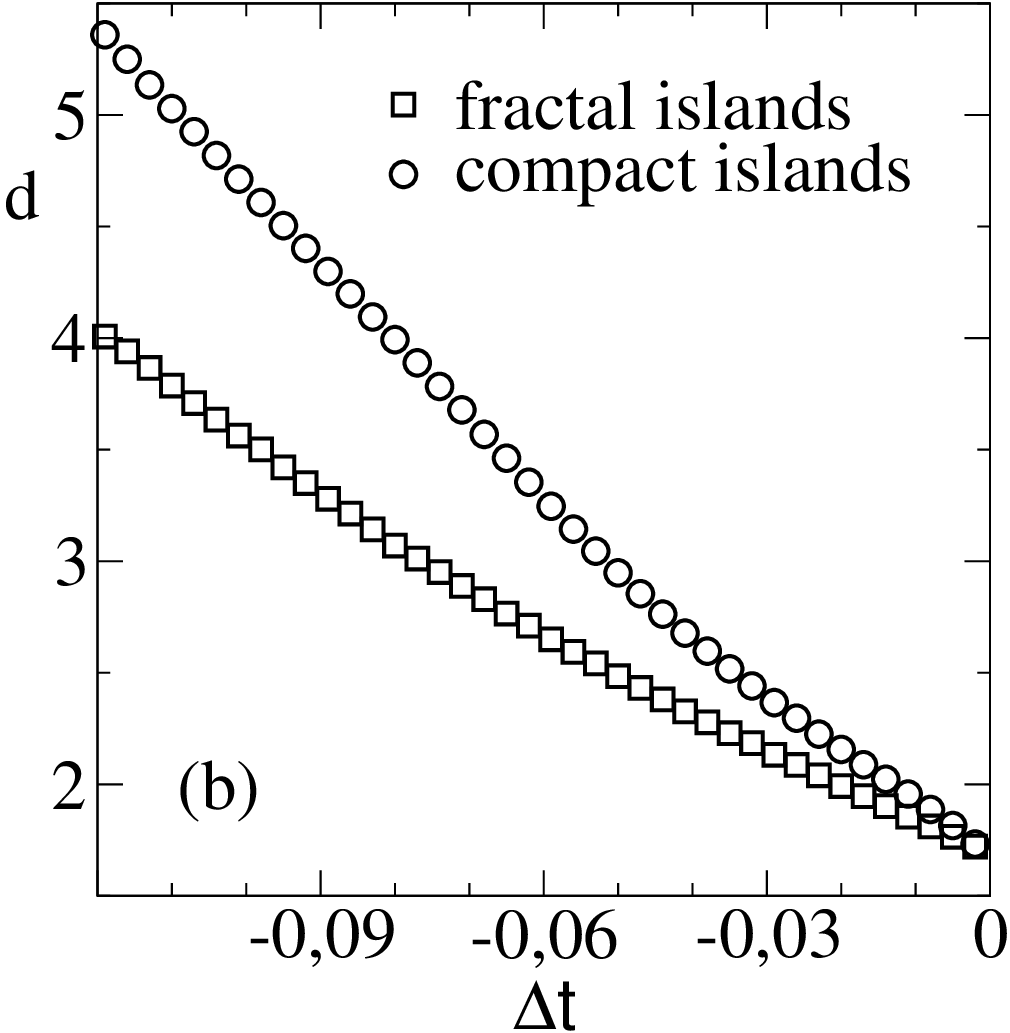}
 \caption{(a) Mean fraction $\Phi$ of distance vectors with
   anti-parallel orientation (with respect to the center of mass
   distance vector), and (b) mean minimal distance $d$ between islands
   as function of the time lag $\Delta t<0$ before a coalescence event
   at zero time. The times are given in units of $F^{-1}$ and the data
   were determined from the KMC simulations for $\Gamma=10^7$.}
 \label{fig:ch5_back_in_time}
\end{figure}

\section{Quantitative analysis of coalescence events}
\label{app:B}
For a quantitative analysis of the coalescence behavior we determined
the fraction of pair distance vectors of coalescing islands that
before coalescence exhibit an anti-parallel orientation to the vector
connecting the center of masses of the islands. Let us denote by
$\mathbf{R}_{i,j}$ the vector pointing from the center of mass of
island $i$ to the center of mass of island $j$, and by
$\mathbf{r}_{i\alpha,j\beta}$ the vector pointing from atom $\alpha$
of island $i$ to atom $\beta$ of island $j$. The fraction of distance
vectors with anti-parallel orientation then is
\begin{equation}
\Phi_{ij}=\frac{1}{s_is_j}\sum_{\alpha,\beta}
H(-\mathbf{r}_{i\alpha,j\beta}\cdot\mathbf{R}_{i,j}) \, ,
\label{eq:f}
\end{equation}
where $H(.)$ is the Heaviside jump function with $H(x)=1$ for $x>0$
and zero else. For a given time lag $\Delta t$ before coalescence, the
$\Phi_{ij}$ were averaged over all coalescence events, yielding the
mean fraction $\Phi(\Delta t)$ of distance vectors with anti-parallel
orientation. To obtain the corresponding data, configurations
generated by the KMC simulations were analyzed afterwards back in
time, starting from the instant where islands first touched each
other.

The mean fraction $\Phi$ obtained from this analysis is shown in
Fig.~\ref{fig:ch5_back_in_time}(a) as a function of $\Delta t$ for
$\Gamma=10^7$.  We assigned negative values to $\Delta t$ to emphasize
that $\Phi(\Delta t)$ was determined for lags before a coalescence
event.  That $\Phi(\Delta t)$ for fractal islands is by many orders of
magnitude larger than for compact islands demonstrates the partial
inter-penetration of the fractal islands before coalescence. The value
$\Phi(\Delta t)\simeq0.01$ reached for the fractal island morphologies
in the limit $\Delta t\to0$ means that on average about 10\% of
the atoms of each island in a coalescence event pass each
other. That the partial inter-penetration is accompanied by a slowing
down of the approach of two islands before coalescence can be seen in
Figure~\ref{fig:ch5_back_in_time}(b), where the averaged minimal
distance $d(\Delta t)$ between coalescing islands is shown, that means
$d_{ij}=\min_{\alpha,\beta}(|\mathbf{r}_{i\alpha,j\beta}|)$ averaged
over all coalescences of islands $i$ and $j$ for time lag $\Delta
t$. The (negative) slope of $d(\Delta t)$ is significantly smaller for
the fractal island morphologies, giving evidence for the screening
effect.\cite{Brune:1998,Brune/etal:1999}


%

\end{document}